%% file: ms.tex
\documentclass[useAMS,usenatbib]{mn2e}
\usepackage{lscape} 
\usepackage{amsmath,amssymb,graphicx,color}
\usepackage{subfigure}

\def\egs{\,erg~s$^{-1}$}      

\def\deg{\hbox{$^\circ$}}

\def\gsim{\;\rlap{\lower 2.5pt\hbox{$\sim$}}\raise 1.5pt\hbox{$>$}\;}
\def\lsim{\;\rlap{\lower 2.5pt\hbox{$\sim$}}\raise 1.5pt\hbox{$<$}\;}
\def\la{\mathrel{\hbox{\rlap{\hbox{\lower4pt\hbox{$\sim$}}}\hbox{$<$}}}}
\def\ga{\mathrel{\hbox{\rlap{\hbox{\lower4pt\hbox{$\sim$}}}\hbox{$>$}}}}

\def\aj{AJ}
\def\apj{ApJ}
\def\mnras{MNRAS}

\title[XMM-Newton Observations of WR binaries]
      {X-ray emission characteristics of two WR binaries : V444 Cyg and CD Cru}
\author[Himali Bhatt et al.]
   {Himali Bhatt$^1$, J. C. Pandey$^1$, Brijesh Kumar$^1$,
    K. P. Singh$^2$, Ram Sagar$^1$  \\
    $^{1}$ Aryabhatta Research Institute of Observational Sciences,
           Manora Peak, Nainital 263 129, India\\
    $^{2}$ Tata Institute of Fundamental Research, Mumbai 400 005, India\\
     }

\begin{document}

\date{\today}


\maketitle

\label{firstpage}

\begin{abstract}

We present X-ray  analysis of two Wolf-Rayet (WR) binaries: V444 Cyg and  CD Cru using the data from  observations with  XMM-Newton. The X-ray light curves show the phase locked variability in both binaries, where the flux increased by a factor of $\sim 2$ in the case of V444 Cyg and $\sim 1.5$ in the case of CD Cru from  minimum to  maximum. The maximum luminosities in the 0.3--7.5 keV energy band were found to be $5.8\times10^{32}$ and $2.8\times10^{32}$ erg s$^{-1}$ for V444 Cyg and CD Cru, respectively. X-ray spectra of these stars confirmed large extinction and revealed hot plasma with prominent emission line features of highly ionized Ne, Mg, Si, S, Ar, Ca and Fe, and are found to be consistent with a two-temperature plasma model.  The cooler plasma at a temperature of $\sim$ 0.6 keV was found to be constant  at all phases of both binaries, and could be due to a distribution of small-scale shocks in radiation-driven outflows.  The hot components in these binaries were found to be phase dependent. They varied from 1.85 to 9.61 keV for V444 Cyg and from 1.63 to 4.27 keV for CD Cru. The absorption of the hard component varied with orbital phase and found to be maximum  during primary eclipse of V444 Cyg.  The high plasma temperature and variability with orbital phase suggest that the hard-component emission is caused by a colliding wind shock between the binary components.

\end{abstract}


\maketitle


\section {Introduction} \label{sec:intro}

 WR stars are known to produce strong stellar winds driven by their strong radiation field.  The stellar winds can reach velocities up to 1000-3000 km s$^{-1}$ with the mass loss rates of $10^{-4}-10^{-6}$ M$_{\odot}$ yr$^{-1}$ depanding upon mass and age (Rauw 2008). These winds not only affect their evolution, but also have a tremendous impact on their surroundings (Rauw 2008).  Most of the O-type and early B-type single stars are reasonably bright ($\rm{10^{31}<L_X< 10^{33}}$ erg s$^{-1}$) and  are soft (kT $< 1$ keV) X-ray sources. The X-ray luminosity  of these sources are found to scale with the bolometric luminosity as ${\rm{L_{\rm X}/L_{\rm bol}}}$ $\sim$ ${10^{-7}}$ (Bergh$\rm{\ddot{o}}$fer et al. 1997, Sana et al. 2006) and is generally thought to be produced due to the shocks, with velocity jumps up to a few hundred km s$^{-1}$, generated throughout the stellar wind due to dynamic instabilities (Lucy \& White 1980; Owocki \& Cohen 1999; Kudritzki \& Puls 2000).  This 'wind-shock' scenario has now become the 'standard' model to explain the X-ray emission from single early-type stars (Rauw 2008 and references therein).  But several massive stars in the Orion Trapezium cluster (Stelzer et al. 2005), M17 cluster (Broos et al. 2007) and  Carina Nebula (Leutenegger \& Kahn 2003) require an additional hard X-ray component to explain their X-ray spectra.  This additional hot component is inconsistent with the standard model which produces only a soft X-rays. 
 In order to explain the hard X-ray component, Babel \& Montmerle (1997) put forward a  magnetically confined wind-shock model. According to this model, the stellar wind is confined by the large-scale magnetic field into the equatorial region and the two streams from the upper and lower  hemispheres collide with it to heat the plasma to temperatures much higher than typically expected for single stars from the standard wind-shock model.However, the  magnetically confined wind-shock model explains hard X-ray on only a small number of single early-type  stars, e.g.,$\theta^1$ Ori C (Gagn$\acute{e}$ et al. 2005) .

  \input{stellar_par.tab}
Stellar binary systems with early type (WR or OB) components are found to
be more luminous in X-ray than what is expected from the individual components separately
(De Becker 2007). This excess X-ray luminosity
is thought to originate from the wind-wind interaction zone where post-shock
temperature is believed to reach up to a few to several tens of  million Kelvin (MK), contributing
mainly to the harder energy ($\rm{kT > 1}$ keV) X-ray emissions. The post-shock
temperature is expected to be higher in long period ($\sim$ a few weeks)
binaries as the stellar winds collide close to their terminal velocities.
In short period ($\sim$ a few days) binaries, the wind collision occurs
while winds are still accelerating  (De Becker 2007). Though, in both the cases, models predict
phase-locked variations in the X-ray domain (Luo, McCray \& Mac Low 1990;
Usov 1992; Antokhin, Owocki \& Brown 2004). These variations are produced
either due to  the varying circumstellar optical depth along the line of sight
towards the shock as the stars revolve around each other or by varying the
orbital separation in eccentric binaries, which changes the intrinsic
strength of the collision (Rauw 2008; Naz$\acute{e}$ et al. 2007).
On the contrary, there are examples of a number of massive star binaries (O+O) which show
a weak enhancement in the X-ray
emission over the single massive stars, but present no variability in the
X-ray flux (Albacete Colombo, M$\acute{e}$ndez \& Morrell, 2003).
 Antokhin (2007) showed that, in general, the X-ray properties of single and
binary early-type stars are not very different.

In this paper, we investigated the X-ray emission characteristics for a two
 short-period  WR binaries, namely V444 Cyg and CD Cru using
high signal-to-noise (S/N) ratio XMM-Newton observations.
For both objects, the X-ray data from XMM-Newton is analysed for the first time,
while for CD Cru,  we analyze the X-ray emission for the first time.
Both the  stars are collected from a systematic study carried out on young galactic clusters using  the XMM-Newton archival data. The choice of these stars associated with young clusters provide us a good knowledge of the distance and  the reddening parameter.
Details of the complete study will be presented elsewhere.
The relevant parameters of the sample are listed in Table~\ref{tab:par_str} and
a description on the individual objects are given below.

V444 Cyg (WR 139) is an O6 III-V + WN5 eclipsing binary system
associated with the cluster Berkeley 86 (van der Hucht 2001) situated at an estimated distance of
1.9 kpc with E(B-V)$\sim$0.8 mag (Massey, Johnson \& Degioia-Eastwood 1995).
The orbital period and an eccentricity of the binary system are estimated to be 4.21 days (d) and 0.03, respectively, (Khaliullin, Khaliullina \& Cherepashchuk 1984).
It has been observed
in multi-wavelengths with well determined physical parameters.
A fairly soft (kT = 0.5 keV) thermal X-ray spectrum was reported
using Einstein observations (Moffat et al. 1982). The flux was found to vary
by a factor of 2 with a minimum at phase zero, which implies that the WN5 star is in front of the O6 star.
Pollock  (1987) reported an
absorption-corrected flux of $\rm {L_ X} = 7.7\times 10^{32}$ erg s$^{-1}$ in
the 0.4-4 keV range. In ROSAT observations, Corcoran et al. (1996) reported
a minimum absorption-corrected X-ray luminosity
of $3.8\times 10^{32}$ erg s$^{-1}$
at the primary eclipse and $8.6\times10^{32}$ erg s$^{-1}$ at the secondary eclipse
 in the 0.4--4.5 keV range. The ASCA observations at three phases in the energy band of 0.7--10 keV were
analyzed by Maeda et al. (1999). They found the presence of
two-temperature (0.6 and 2.0 keV) plasma, with soft component (0.2--4 keV)
luminosity of $6-11 \times 10^{32}$ erg s$^{-1}$.
 The high temperature plasma (kT $\sim$ 2 keV) and the variability with the orbital phase
 suggest that the hard component emission
is caused due to the shocks originated from the collision of winds from WN5 and O6 stars (Maeda et al. 1999).

CD Cru (WR 47) is an O5 V + WN6 binary system reported as a member of the cluster Hogg 15
(van der Hucht 2001). Throughout this study, we adopted a distance of 3 kpc of Hogg 15 (see Piatti et al. 2002 ;Sagar, Munari \& Boer 2001). The orbital period of this binary system is 6.24 d with zero eccentricity (Niemela, Massey \& Conti 1980).
No previous X-ray observations of the system are reported in the literature.

Our paper is organised in the following manner:
in Section 2 and 3, we give a detailed description of  X-ray data reduction and analysis, respectively.
In Section 4, we present a discussion on the X-ray properties of the sources.
 Finally, we summarize and draw our conclusions in Section 5.

  \input{xray_observations.tab}

\section{X-ray Observations and Data Reduction}
\label{sec:reduc}
The log of the X-ray observations is given in Table~\ref{tab:xray_observations}.
We made use of the archival data obtained with the XMM-Newton observatory
which consists of three co-aligned X-ray telescopes observed simultaneously
and covered 30$^\prime$ $\times$ 30$^\prime$ region of the sky. The X-ray
photons were recorded with the European Photon Imaging Camera (EPIC), which
forms images on three CCD-based detectors: the PN (Str$\ddot{u}$der et al. 2001),
and the twin MOS1 and MOS2 (Turner et al. 2001) with an angular resolution
of 6$^{\prime\prime}$ (FWHM). During the observations, all the three
EPIC detectors were active in full frame mode together with the Thick filter.
For V444 Cyg, six separate observations were taken with exposure time ranging from 10 to 20 ks. These observations are spread over the half cycle of binary system. However, for the star CD Cru, three separate observations were taken covering almost full binary cycle.

\subsection{EPIC data reduction}
\label{sec:src_reduc}

We reduced the X-ray data using standard XMM-Newton Science Analysis System
software (SAS version 7.0.0) with updated calibration files (Ehle et al. 2004).
Event files for the MOS and the PN detectors were generated  using the tasks
{\sc emchain} and {\sc epchain}, respectively. These tasks allow calibration of the
energy and the astrometry of the events registered in each CCD chip and to
combine them in a single data file. We restricted our analysis  to the energy band
0.3--7.5 keV, as the data below 0.3 keV are mostly unrelated to bonafide
X-rays, and above 7.5 keV is mostly dominated by the background counts.
Event list files were extracted using the SAS task {\sc evselect}.
Data from the three cameras were individually screened for the time intervals
with high background when the total count rates (for single
events of energy above 10 keV) in the instruments exceed 0.35 and 1.0
$\rm{counts~s^{-1}}$ for the MOS and PN detectors, respectively.
The observations with observation ID 0206240401 for the source V444 Cyg were heavily affected
by the high background events, and the data in MOS detectors were not useful.
Only half of the observation time ($\sim$5 ks)
of PN was found useful.

Light curves and spectra were extracted using a circular region
with the source as the center in the energy range 0.3--7.5 keV of the EPIC detectors.
The X-ray sources in the cluster were often found to be largely contaminated due to the emission from
the neighboring sources. For this reason, the radii of extraction regions were varied
between 20$^{\prime\prime}$ and 30$^{\prime\prime}$ depending upon the position
of the sources in the detector and their angular separation between the
neighboring X-ray sources. The background has been estimated from
a number of empty regions close to the X-ray source in the same CCD of the detector.
X-ray spectra of the sources were generated using SAS task {\sc especget}, which
also computes the photon redistribution as well as the ancillary matrix.
For each source, the background spectrum was obtained from
regions devoid of any sources chosen according to the source location. Finally, the spectra were
re-binned to have at least 20 counts per spectral bin for both the sources in all the observations.

\subsection{RGS data reduction}
\label{sec:rgs}
The Reflection Grating Spectrometers (RGS) are mounted on two of the
three XMM-Newton X-ray telescopes and were operated in spectroscopy
mode during the observations. We followed the standard procedure
as outlined in the XMM-Newton handbook to generate the RGS spectra of
the sources. The raw data were processed with the task {\sc rgsproc}
at the position of the sources.
The other bright sources, found in the same field observed by RGS, have been
excluded from background estimation. In addition, we have filtered the
event list for high background level events. For star V444 Cyg,
the data obtained in observations with observation ID 0206240401 were contaminated with
high background events in RGS detector, therefore the data was not useful.
For CD Cru, the two observations ID 0109480101 and 0109480201
were not useful as they are heavily contaminated by a brighter source HD 110432
and affected by high background episodes, respectively.

\section {Analysis and Results}\label{sec:results}

\subsection{X-ray light curves}\label{sec:WR_lt}
\begin{figure*}
\centering
\subfigure[]{\includegraphics[width=3.0in]{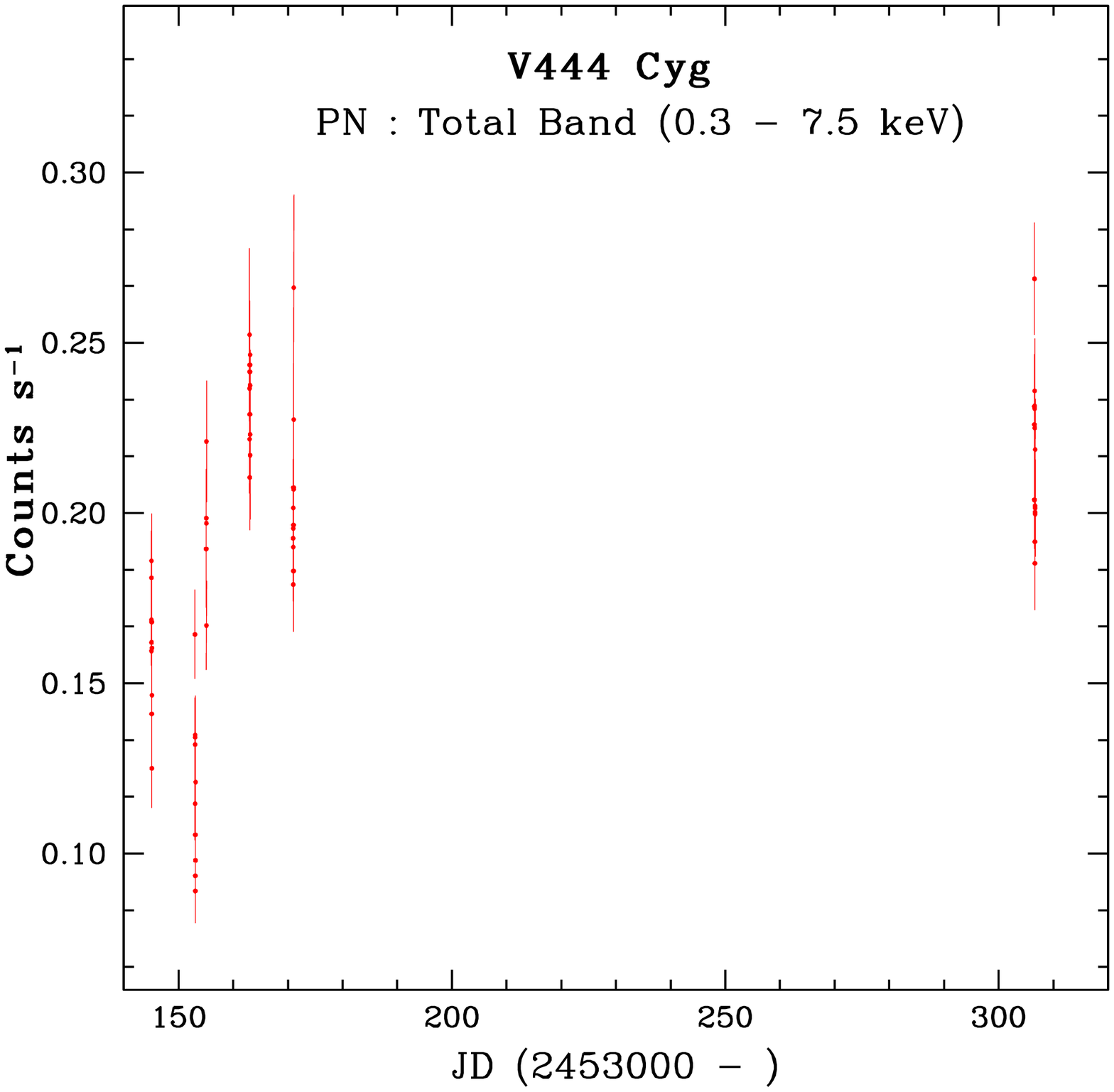}}
\subfigure[]{\includegraphics[width=3.0in]{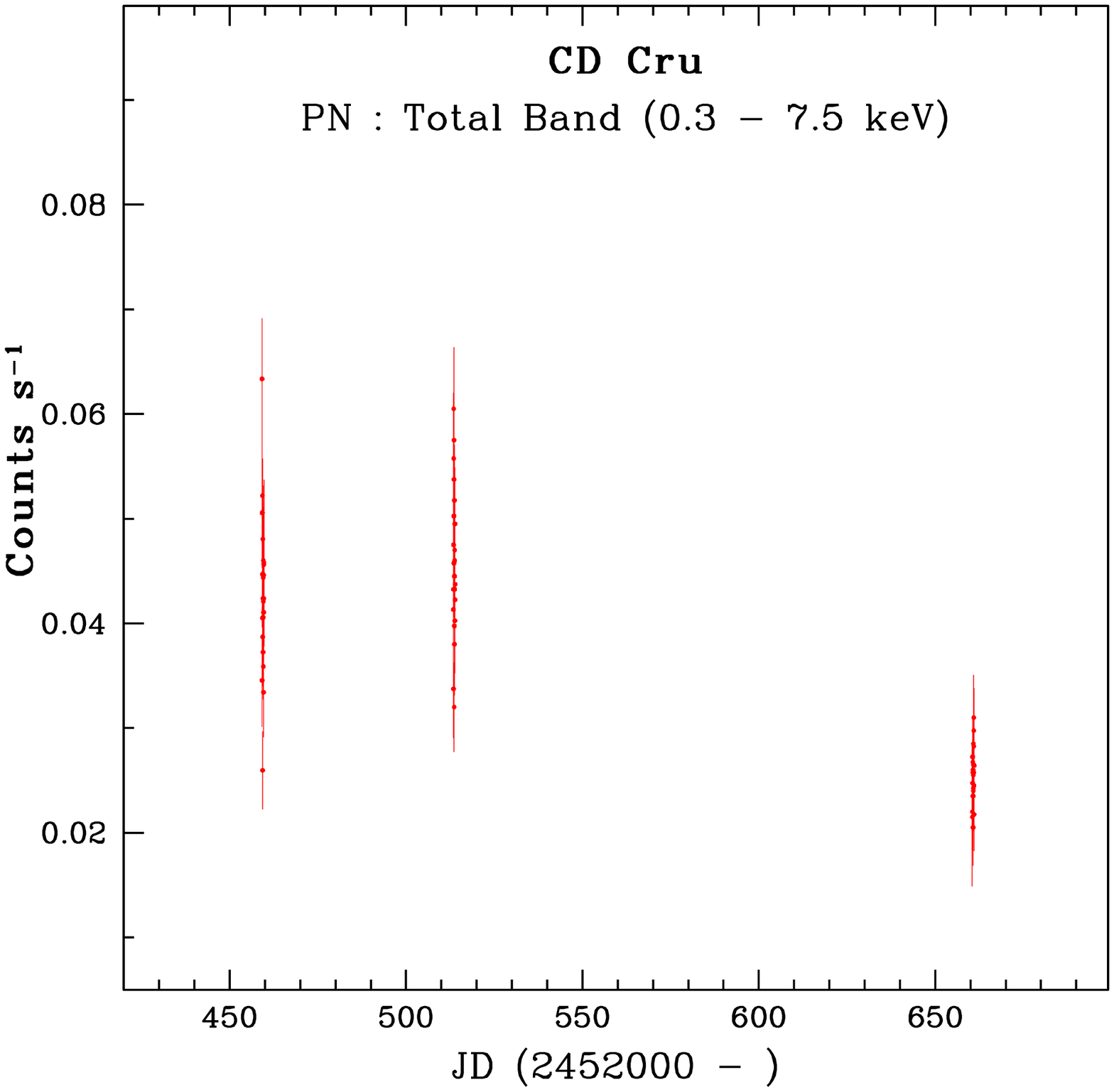}}
\caption{X-ray light curves of the WR-binaries (a) V444 Cyg and (b) CD Cru in the 0.3--7.5 keV energy band.}
\label{fig:xlc}
\end{figure*}

\begin{figure*}
\centering
\subfigure[]{\includegraphics[width=3.0in]{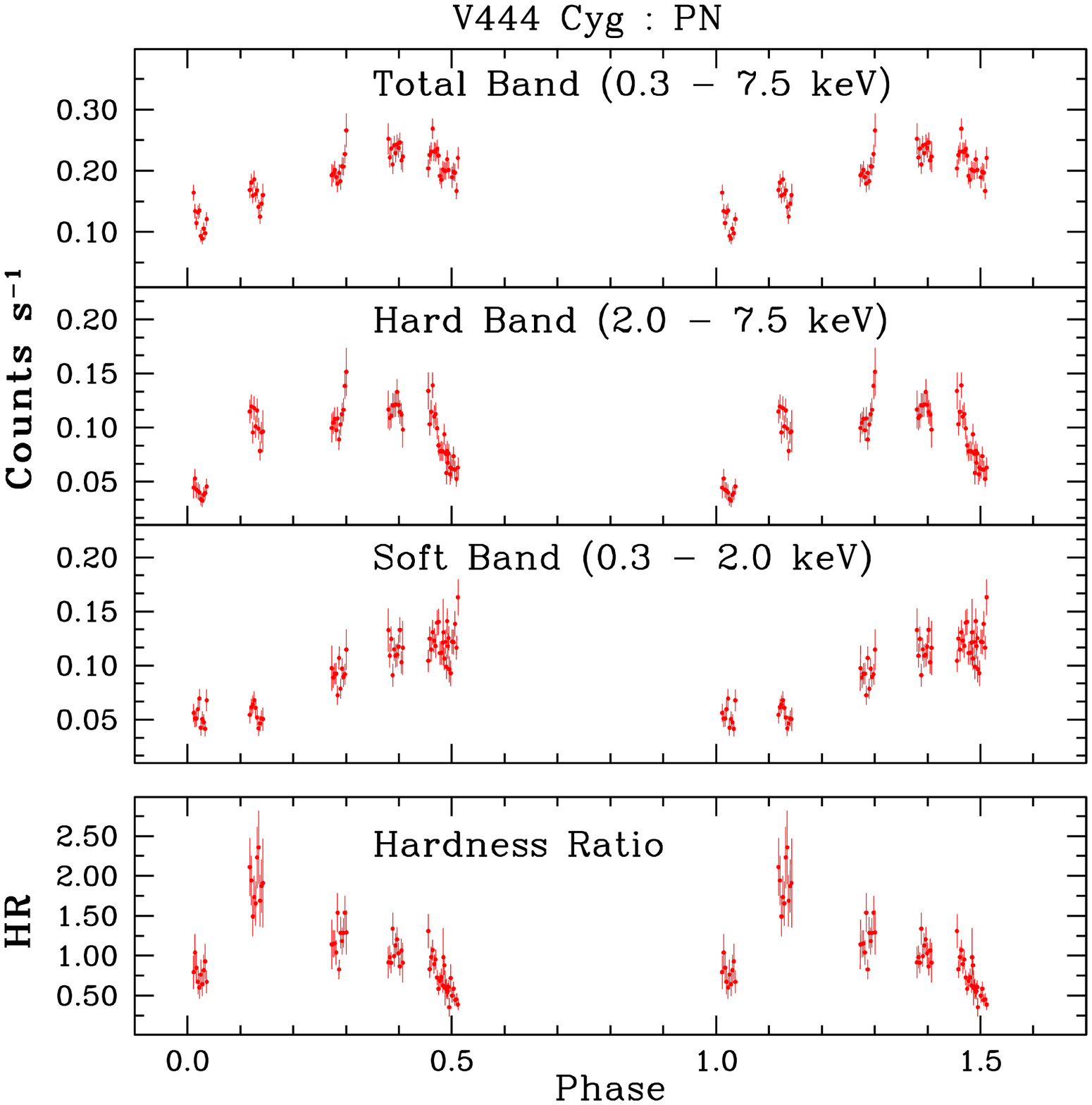}}
\subfigure[]{\includegraphics[width=3.0in]{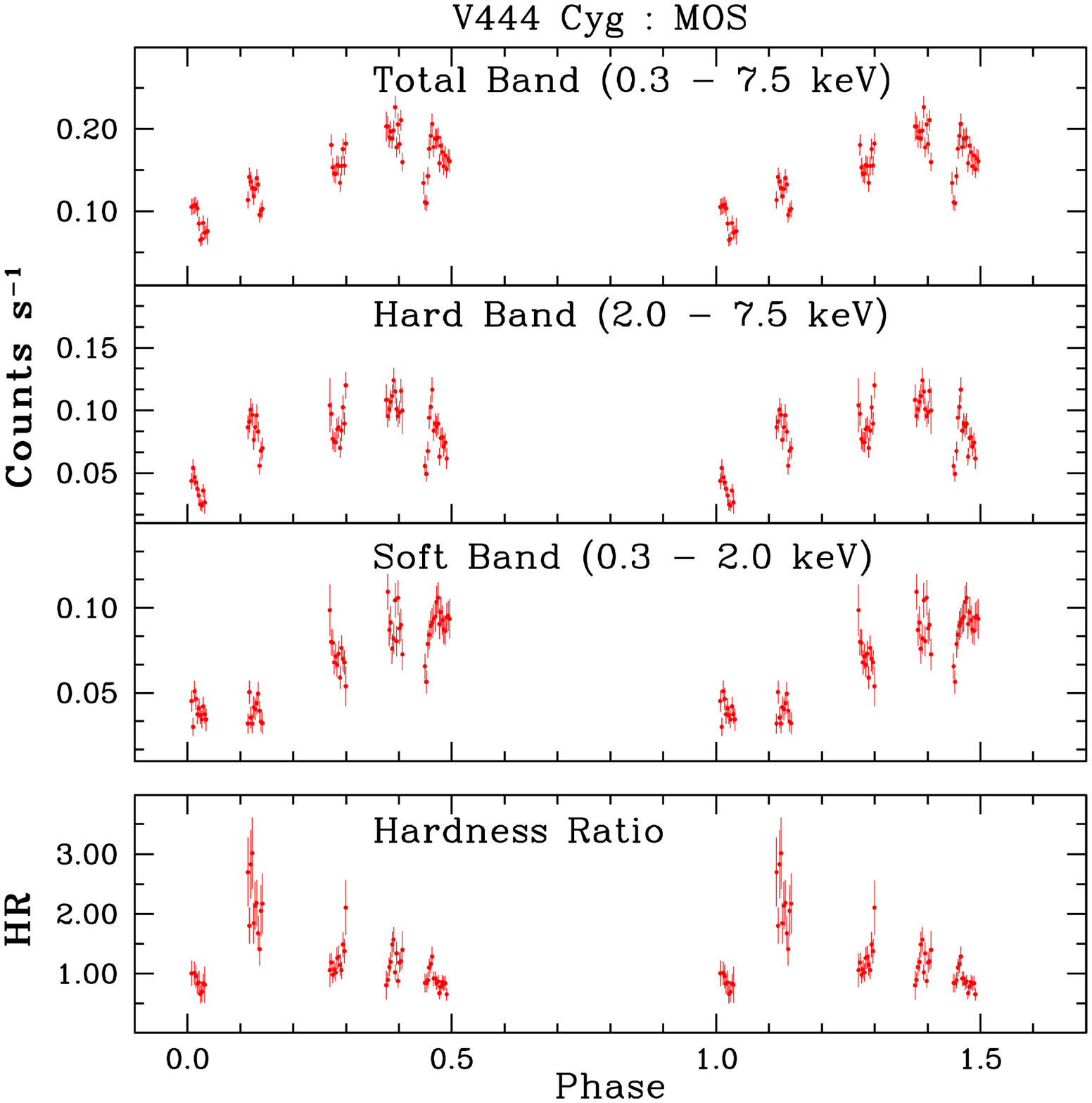}}
\caption{(a) PN and (b) MOS light curves at three energy bands: the total (0.3--7.5 keV), the soft (0.3--2.0 keV) and the hard (2.0--7.5 keV), and hardness ratio  HR curve as a function of orbital phase, where HR = hard/soft of V444 Cyg.}
\label{fig:V444Cygphaselc}
\end{figure*}

\begin{figure*}
\centering
\subfigure[]{\includegraphics[width=3.0in]{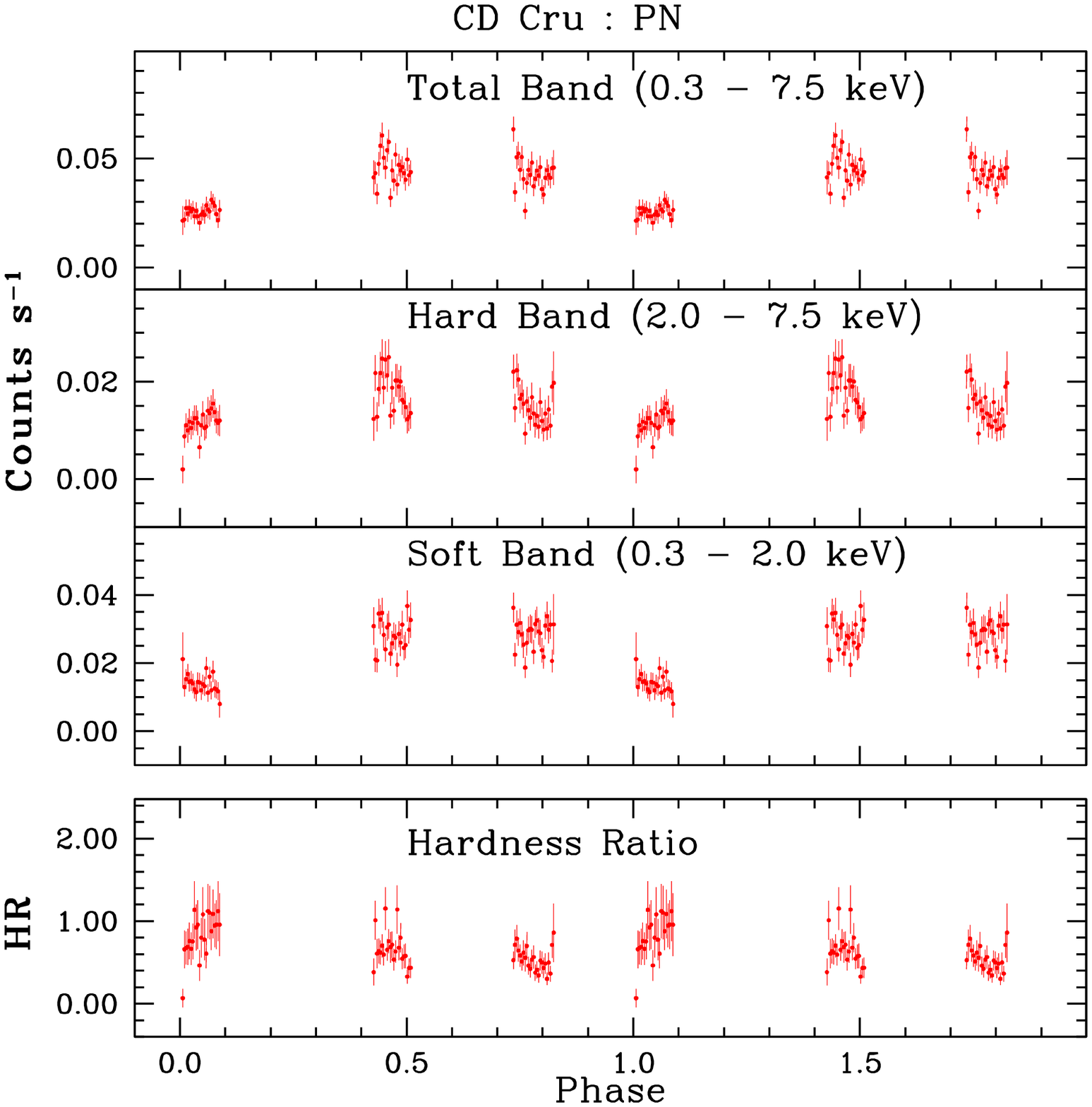}}
\subfigure[]{\includegraphics[width=3.0in]{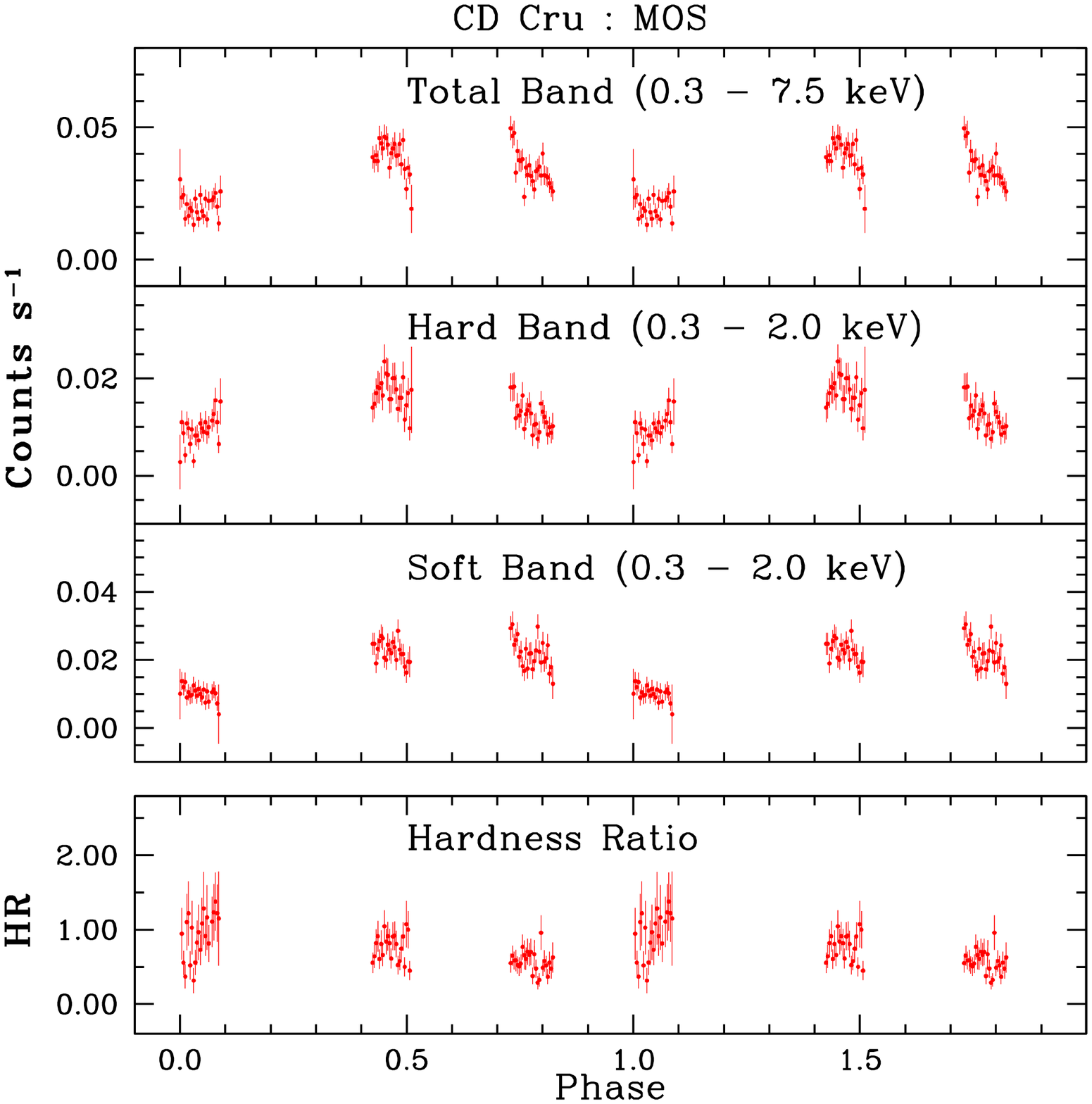}}
\caption{Similar to Fig. \ref{fig:V444Cygphaselc} but for CD Cru.}
\label{fig:CDCruphaselc}
\end{figure*}

The background subtracted X-ray light curves of the WR-binaries V444 Cyg and CD Cru, as observed with PN detector are shown in Fig. \ref{fig:xlc}(a) and \ref{fig:xlc}(b), respectively. The light curves are in the  0.3--7.5 keV (total) energy band show the variability which suggests colliding wind shocks.
Further, we performed the $\rm{\chi^2}$ test  to measure
the significance of the deviations from the mean count rate in order to quantify the constancy of
the data over the time-scale of observations.
We found the variability in the light curves with a confidence level of greater than 99.999\% for both binaries. In order to investigate the variability in the different energy bands, the light curve of these binaries obtained with the MOS and PN data are  generated into  two energy bands namely the soft (0.3--2.0 keV) and the hard (2.0--7.5 keV). The hardness ratio (HR) is defined by the ratio of hard to soft band count rates. The total, hard and soft band intensity curves, and the HR curve  as a function of the orbital phase are shown in the subpanels running from top to bottom in  Fig. \ref{fig:V444Cygphaselc}a  (PN) and \ref{fig:V444Cygphaselc}b (MOS) for the WR binary V444 Cyg. Similar plots of intensity and HR for the star CD Cru are shown in Fig. \ref{fig:CDCruphaselc}. The phases of the observations are reckoned using the ephemeris HJD = 2441164.337+4.213E for V444 Cyg (Underhill, Grieve \& Louth  1990) and  HJD = 2443918.4000+6.2399E  for  CD Cru (Moffat et al. 1990, Niemela, Massey \& Conti 1980). Here, the phase 0.0 indicates the primary eclipse and the phase 0.5 indicates the secondary eclipse. The light curves in the individual bands show the phase locked variability for both binaries.  For V444 Cyg, the count rates in total energy band were minimum at phase 0.0 and maximum at the phase 0.45. After the phase 0.45, the count rates were decreased upto the phase 0.5. The count rates are increased by a factor of $\sim 2$ being minimum at phase 0.0\ in  the total energy band.  In the soft band, the count rates were minimum at phase 0.0 and maximum during the phase 0.45-0.5.  However, in the hard band light curve two minima at phase 0.0 and 0.5 were seen clearly. During the phases from 0.13 to 0.45 (i.e. outside the eclipse)  the count rates in the hard band were constant. The HR curve can reveal the information about the spectral variations.  The HR curve  shows that the emitted X-ray  is harder just after phase 0.0 being maximum at phase 0.13, and afterward decreased till the phase 0.5.
In the case of CD Cru, the nature of the variability was found to be similar to that seen in V444 Cyg.  In soft band, the count rates were found to be constant during the phases 0.47 and 0.78 (i.e. outside the eclipse). However,  the count rates are decreased rapidly after the phase 0.47 \ in the hard band.  A small variation was also observed in the HR curve of the CD Cru. It was maximum at the phase 0.0 and minimum during  the phase 0.47.

\subsection{X-ray spectra and spectral fits }\label{sec:WR_spt}
\begin{figure*}
  \centerline{\hbox{\hspace{0.5in}
\includegraphics[width=60mm, angle=-90]{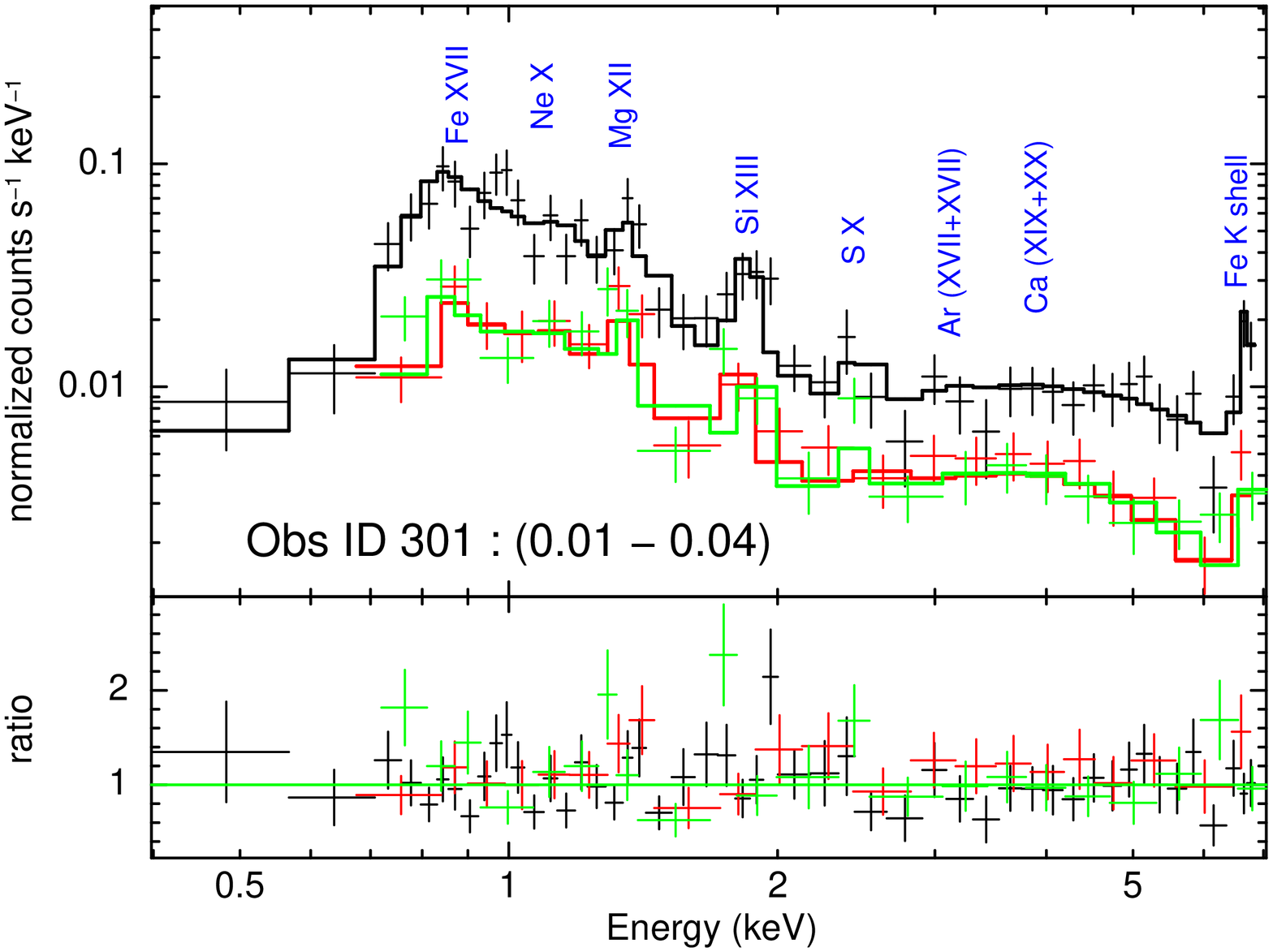}
    \hspace{0.15in}
\includegraphics[width=60mm, angle=-90]{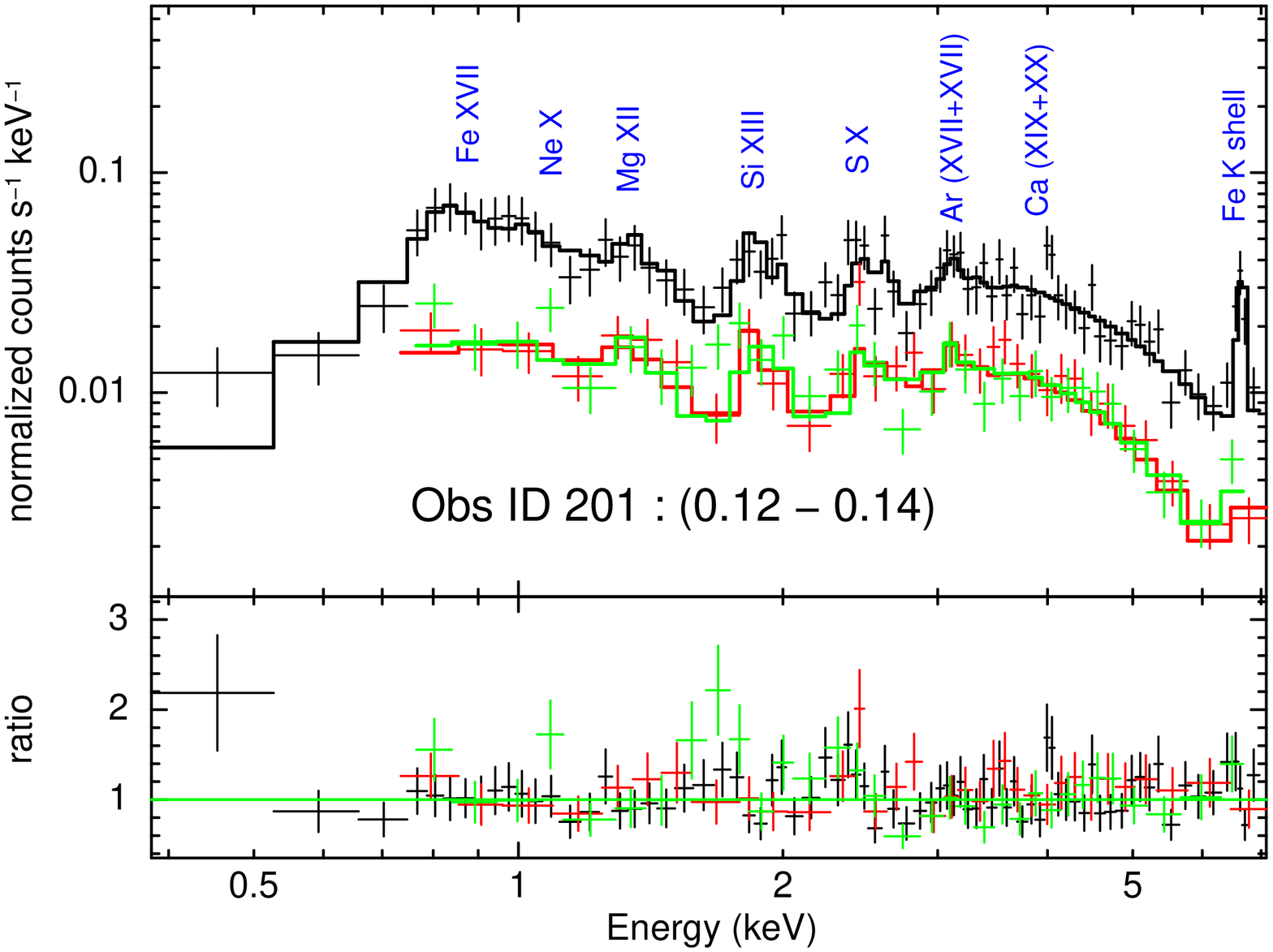}
}}
    \vspace{0.15in}
\centerline{\hbox{\hspace{0.5in}
\includegraphics[width=60mm, angle=-90]{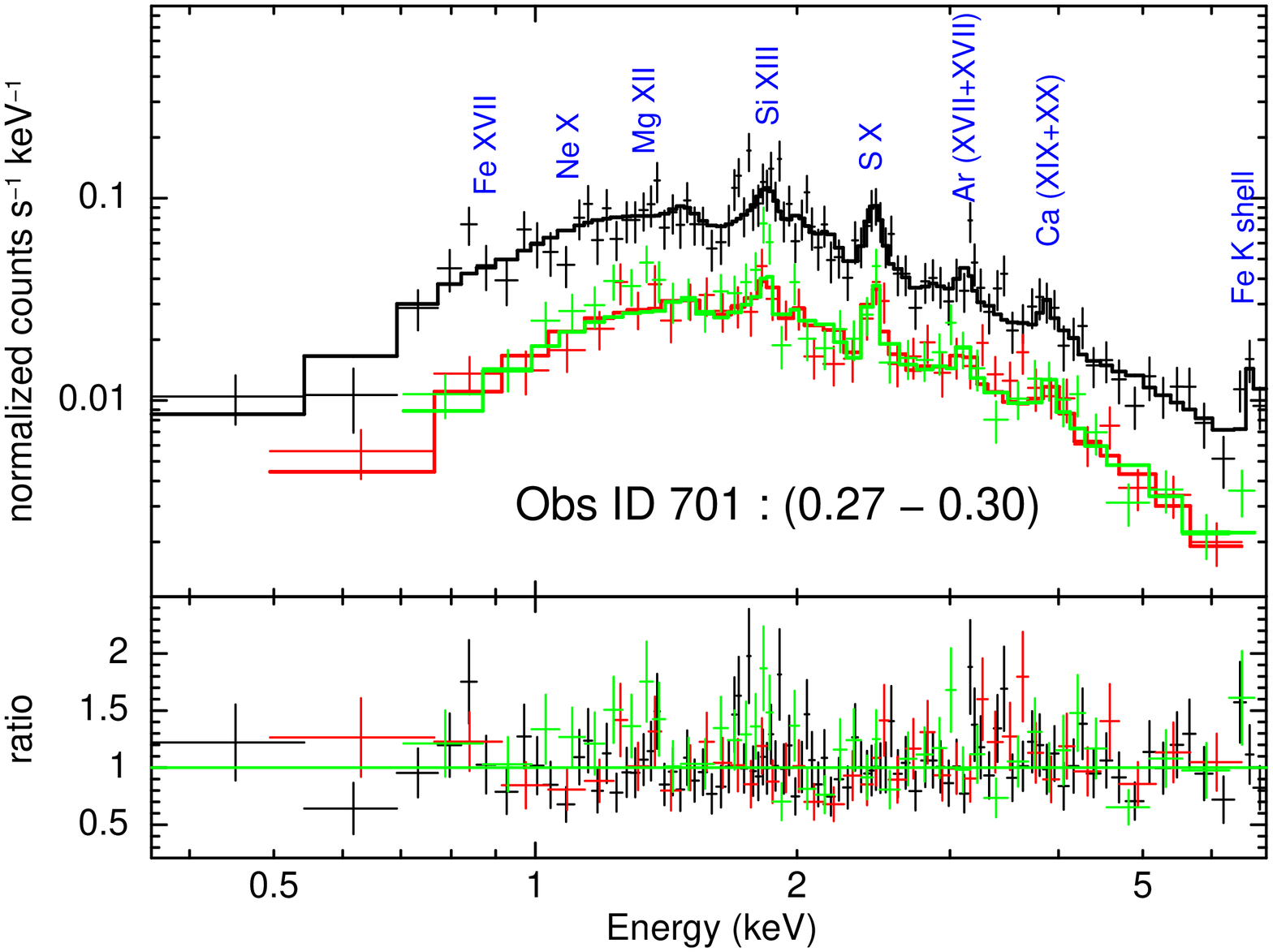}
    \hspace{0.15in}
\includegraphics[width=60mm, angle=-90]{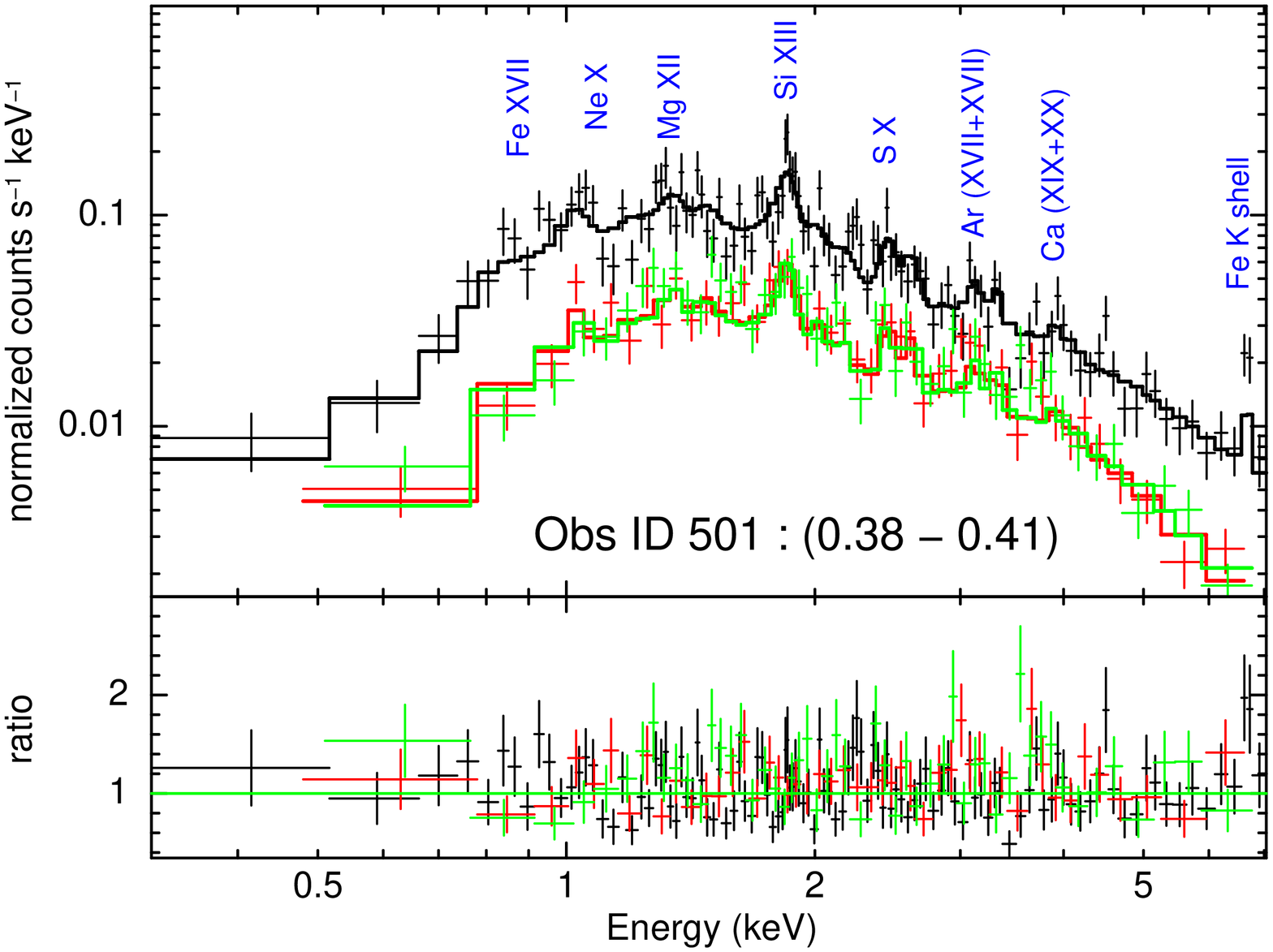}
}}
    \vspace{0.15in}
\centerline{\hbox{\hspace{0.5in}
\includegraphics[width=60mm, angle=-90]{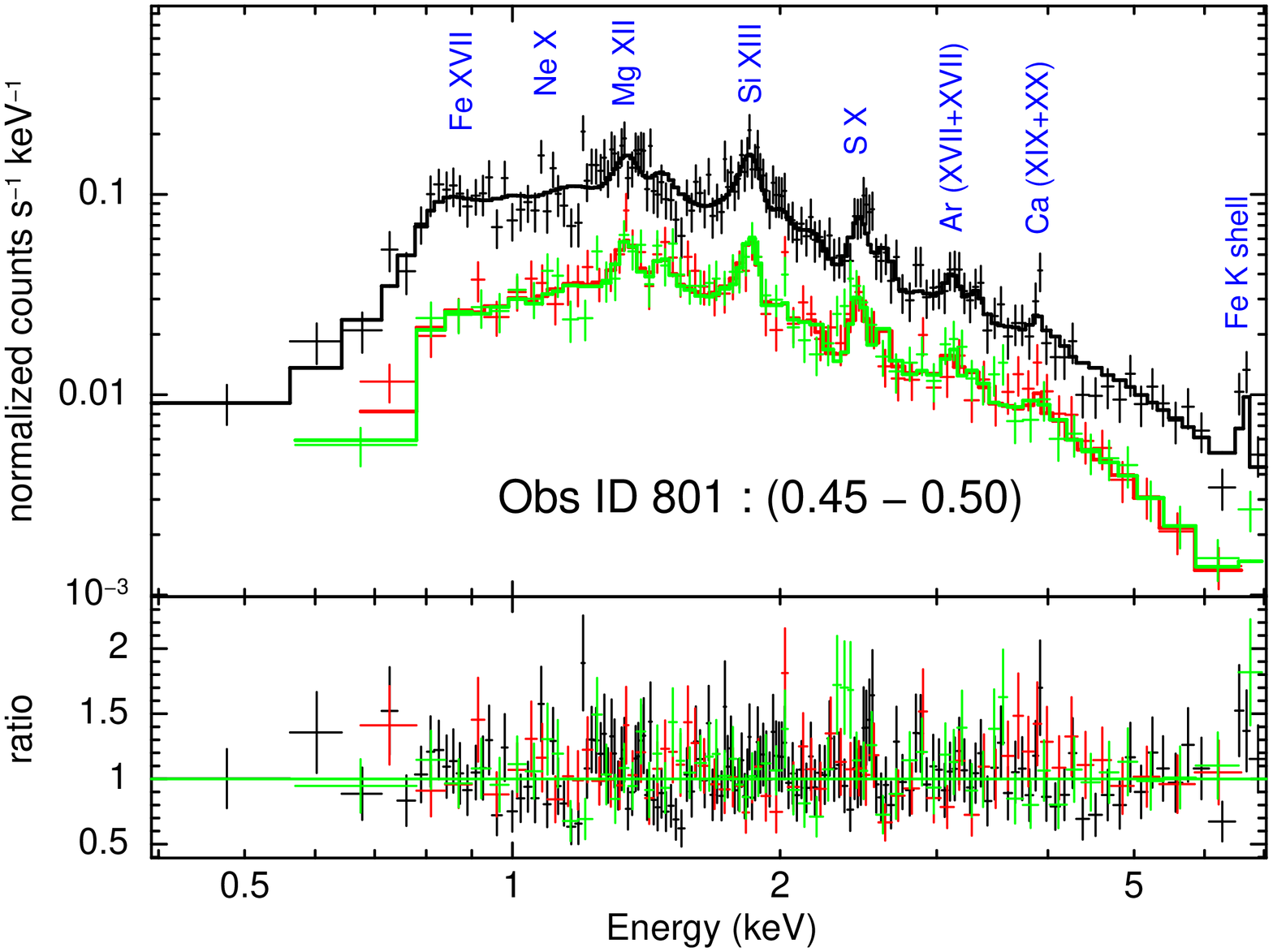}
    \hspace{0.15in}
\includegraphics[width=60mm, angle=-90]{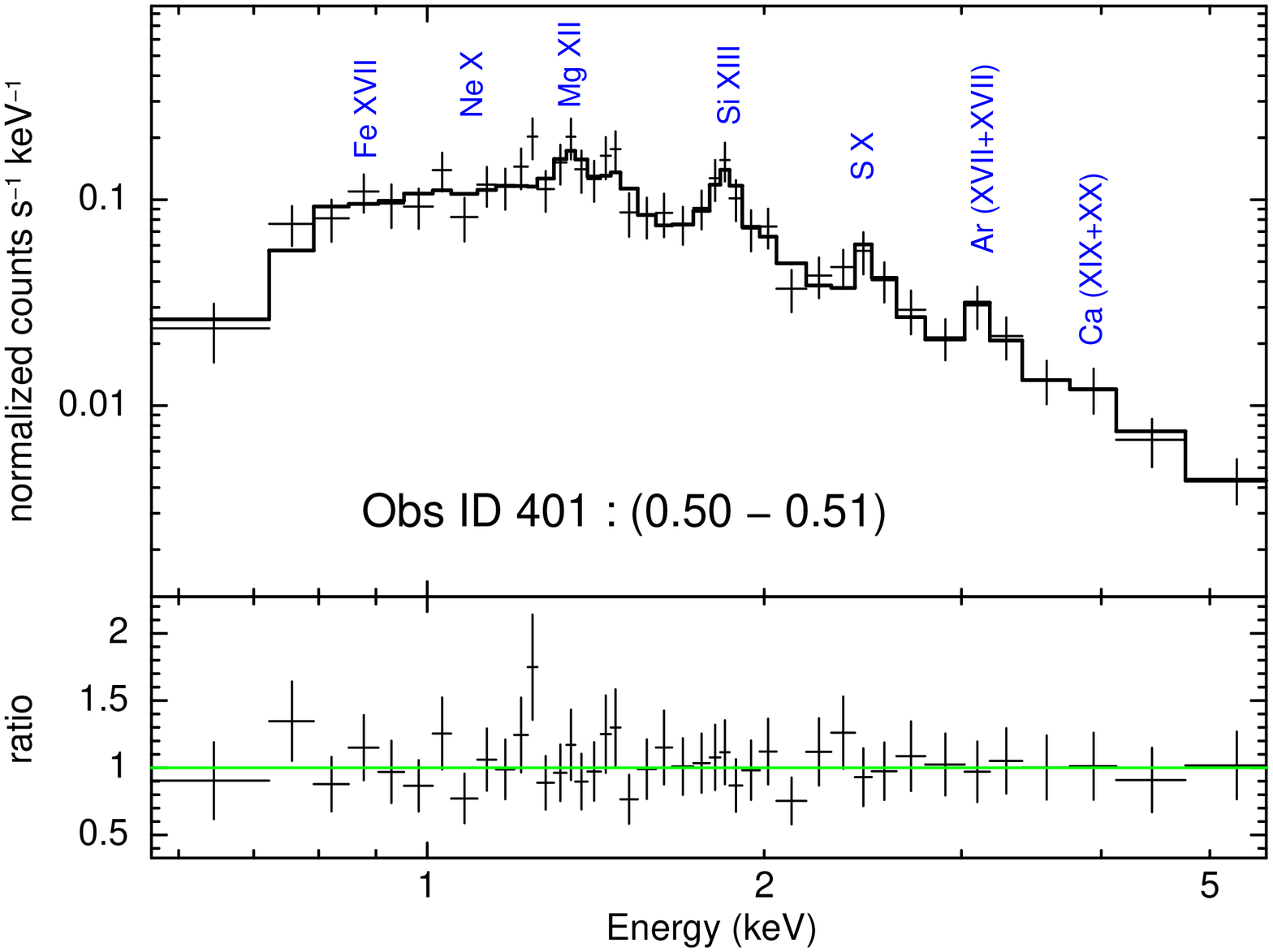}
}}
\vspace{3pt}
\caption{X-ray  spectra of MOS and PN data with the best fit 2T {\sc vapec}
model in upper subpanels of each graph for V444 Cyg. The $\chi^2$ distribution in terms of ratio are given in lower subpanels of each graph. The last three digits of observation ID and the corresponding phases of the observations are given in each graph.
  }
\label{fig:spt_V444Cyg}
\end{figure*}

\begin{figure*}
  \centerline{\hbox{\hspace{0.5in}
\includegraphics[width=60mm, angle=-90]{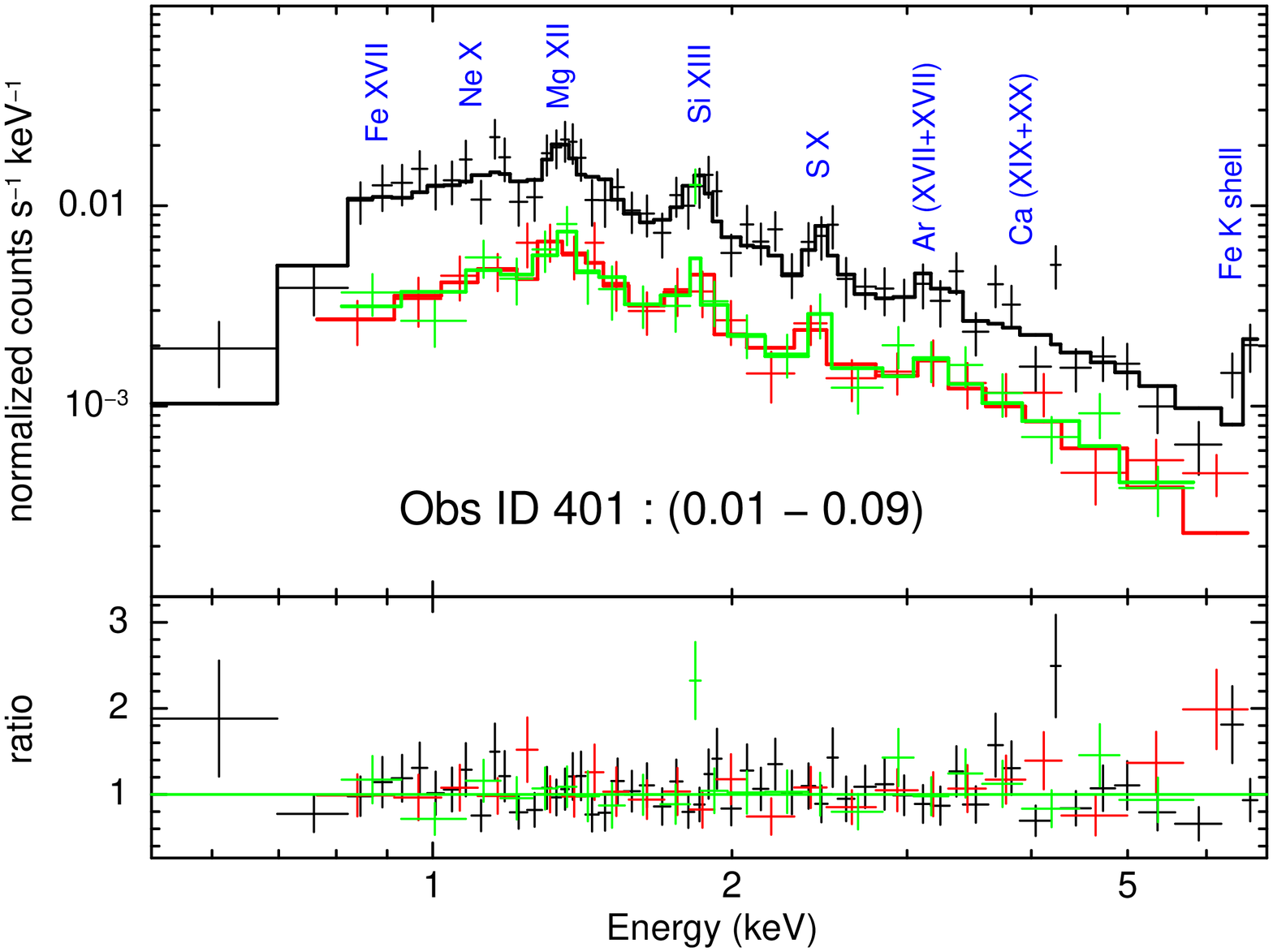}
    \hspace{0.15in}
\includegraphics[width=60mm, angle=-90]{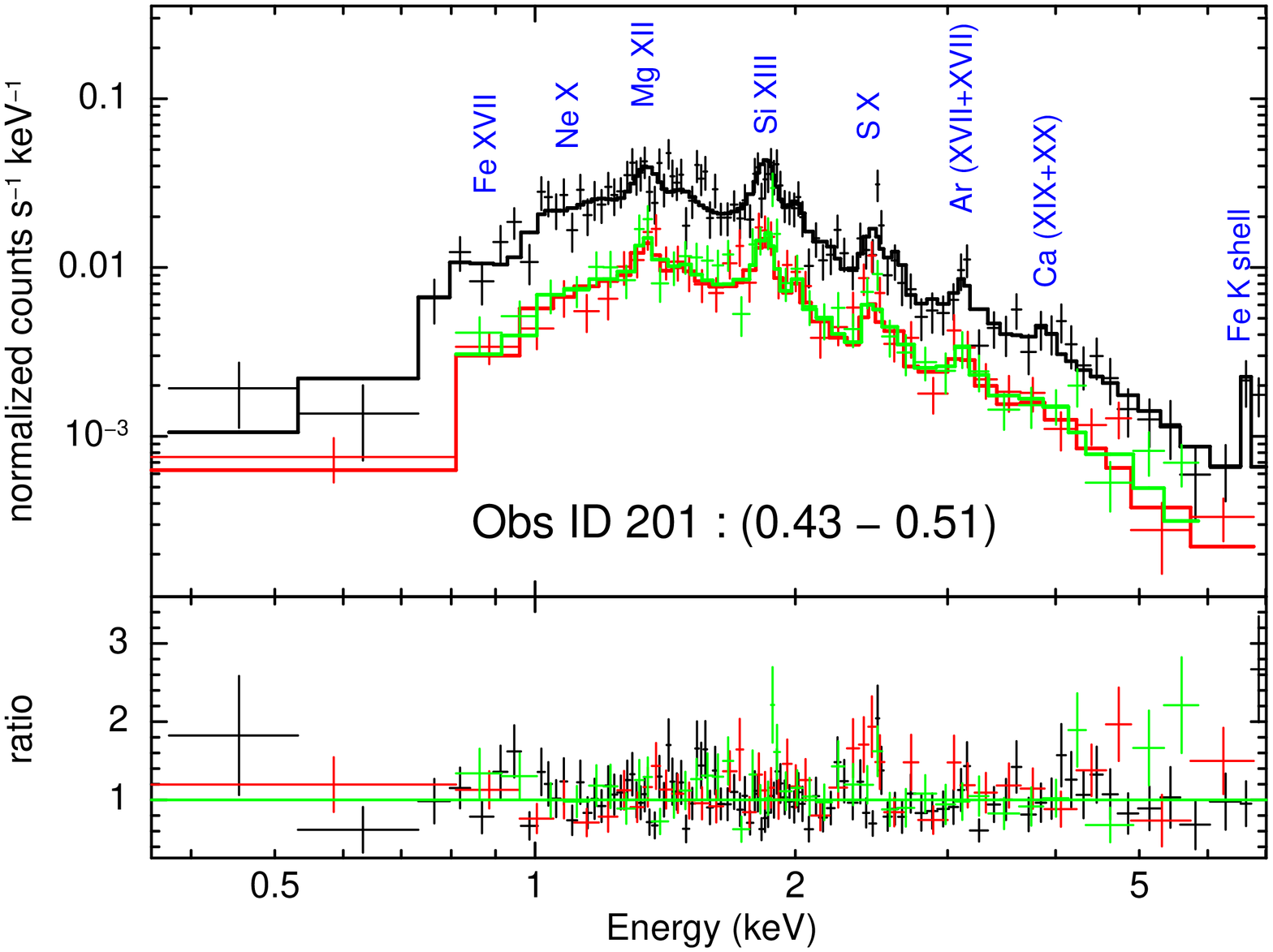}
}}
    \vspace{0.15in}
\centerline{\hbox{\hspace{0.5in}
\includegraphics[width=60mm, angle=-90]{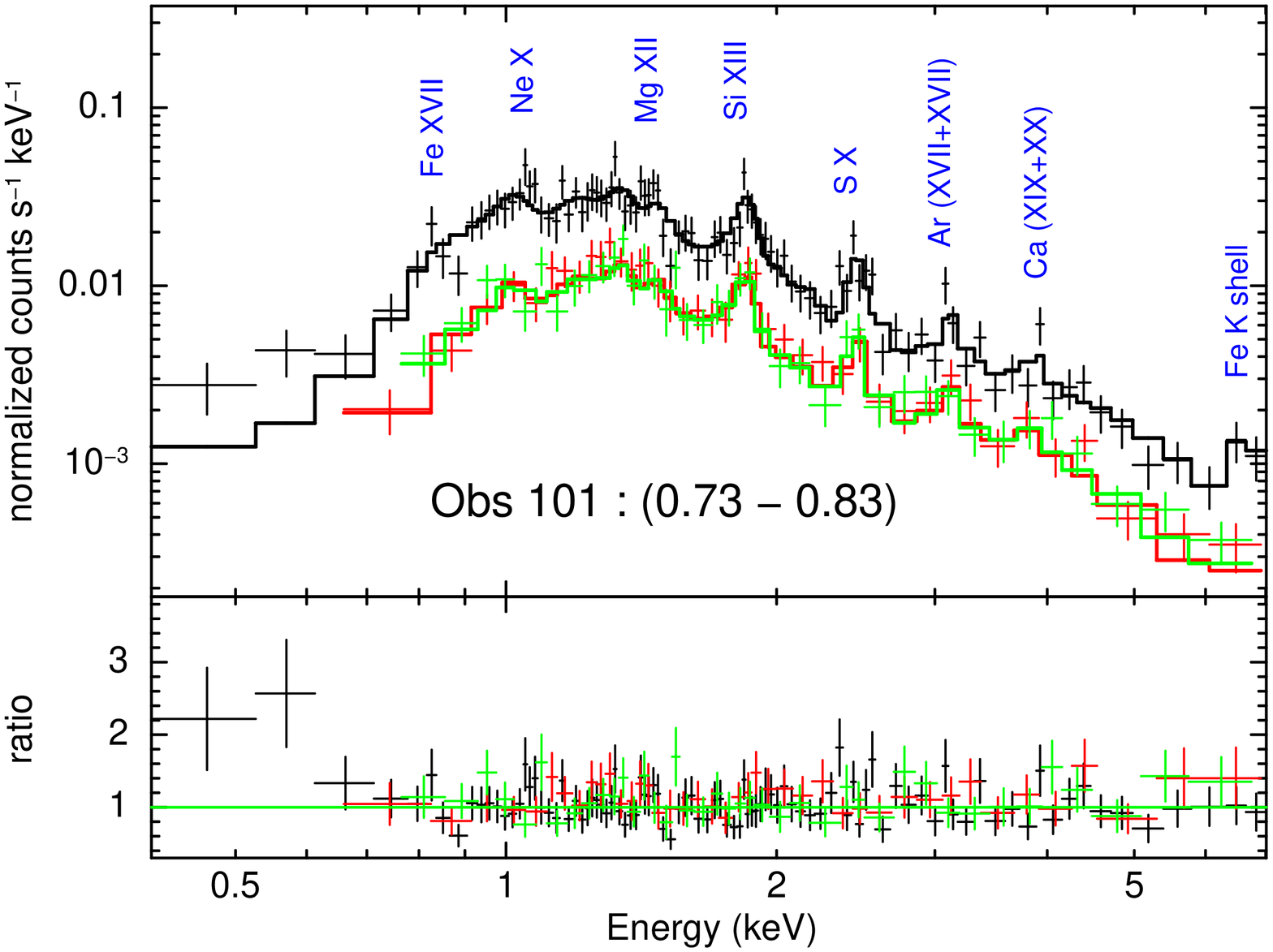}
}}
\vspace{3pt}
\caption{Same as Figure~\ref{fig:spt_V444Cyg}  but for CD Cru.
  }
\label{fig:spt_CDCru}
\end{figure*}

The EPIC spectra of WR binaries in different phases are shown in
Figs.~\ref{fig:spt_V444Cyg} and \ref{fig:spt_CDCru} for V444 Cyg and CD Cru, respectively.
Below 1 keV, the spectra were found to be affected by the high extinction as previously
observed with ASCA for V444 Cyg (Maeda et al. 1999).
Strong emission lines are seen in the MOS and PN spectra of both WR binaries.
 The most prominent lines found in the spectra along with their laboratory
energies are the following: Fe XVII (0.8 keV), Ne X (1.02 keV), Mg XII (1.47 keV),
Si XIII (1.853 keV), S XV (2.45 keV), Ar XVII (3.12 keV), Ca XIX+XX (3.9 keV)
and Fe XXV (6.63 keV).

In order to trace the spectral parameters at different binary phases, we performed spectral analysis of each data set using simultaneous fitting of EPIC data
by two models, (a) plane-parallel shock model ({\sc vpshock}; Borkowski, Lyerly \& Reynolds 2001),
and (b) models of Astrophysical Plasma Emission Code
({\sc apec}; Smith et al. 2001), as implemented in the XSPEC version 12.3.0.
A $\chi^2$ -- minimization  gave the best fitted model to the data.
The presence of interstellar material along the line-of-sight and the local
circumstellar material around the stars can modify the X-ray emission from massive stars.
We corrected for the local  absorption in the line-of-sight to the source using
the photoelectric absorption cross
sections according to Baluci$\acute{n}$ska-Church \& McCammon (1992) and
modeled as {\sc phabs} (photoelectric absorption screens) with two absorption
components, $\rm{N_{H}^{ISM}}$ and $\rm{N_{H}^{local}}$.
The $\rm{N_{H}^{ISM}}$ was estimated using the relation,
$\rm {N_H}$ $\rm{= 5.0\times}10^{21}\times{E(B-V)~cm^{-2}}$ (Vuong et al. 2003), where
$\rm{E(B-V)=A_V/3.1}$, assuming a normal interstellar reddening law towards the direction
of the cluster. We used the values of $\rm{A_V}$ nearly 2.48 mag and 3.56 mag derived for the cluster Berkeley 86
(Massey, Johnson \& Degioia-Eastwood 1995) and Hogg 15 (Sagar, Munari \& Boer 2001), respectively.
The estimated values of $\rm{N_{H}^{ISM}}$ towards V444 Cyg and CD Cru are found to be
$\rm{4.0\times10^{21}~cm^{-2}}$ and $\rm{6.0\times10^{21}~cm^{-2}}$, respectively.
The $\rm{N_{H}^{local}}$ was estimated by making a fit to the observed spectra by varying
the local environment for
the soft ($\rm{kT_1}$) and the hard ($\rm{kT_2}$) energy components in terms of $\rm{N_H^1}$ and $\rm{N_H^2}$, respectively.
Because the WN stars are at  evolved stages, the abundances of He, C, N, O, Ne, Mg, Si, S,
Ar, Ca, and Fe were allowed to vary during the fitting procedure to account
for the observed line emission. The solar abundances were adopted from
Lodders (2003).

First, we fitted {\sc vpshock} plasma model to derive their
spectral features. The constant temperature {\sc vpshock} plasma
model was considered without incorporating mass-loss and orbital parameters of WR stars.
However, the model does account for
non-equilibrium ionization effects and assumes an equal
electron and ion temperature. The best fit {\sc vpshock} models to the
data for the WR stars  are given in Table~\ref{tab:V444Cyg_fit} and Table~\ref{tab:CDCru_fit}
for V444 Cyg and CD Cru, respectively.

X-ray emitting plasma may not be isothermal and
the observed X-ray spectrum may be a superposition of a cool stellar component
and a hot colliding wind shock plasma, as a number of emission lines in the spectra can be formed
over a range of temperatures.
The cooler component is believed to arise from the instabilities
in radiation-driven outflows. However, using the X-ray imaging data of
XMM-Newton it is not possible to resolve the X-ray emission from colliding
wind region and individual stars separately. Therefore, secondly, we fitted two
temperature (2T) plasma model "{\sc vapec}" to characterize such components. The form of 2T plasma model was {\sc phabs(phabs*vapec+phabs*vapec)}.
 In terms of $\chi^2$ the 2T plasma model  provides  better goodness-of-fit for both the sources over the {\sc vpshock} model.
The results for the 2T plasma model  along with the data are displayed
in Fig.~\ref{fig:spt_V444Cyg} and Fig.~\ref{fig:spt_CDCru}  for
V444 Cyg and CD Cru, respectively. The best-fit parameters
are given in Table~\ref{tab:V444Cyg_fit} and  Table~\ref{tab:CDCru_fit}
for V444 Cyg and CD Cru, respectively.
  
  \input{bestfit_WR_V444Cyg.tab}
  \input{bestfit_WR_CDCru.tab}

\clearpage

\begin{figure}
\centering
\subfigure[]{\includegraphics[width=3.0in]{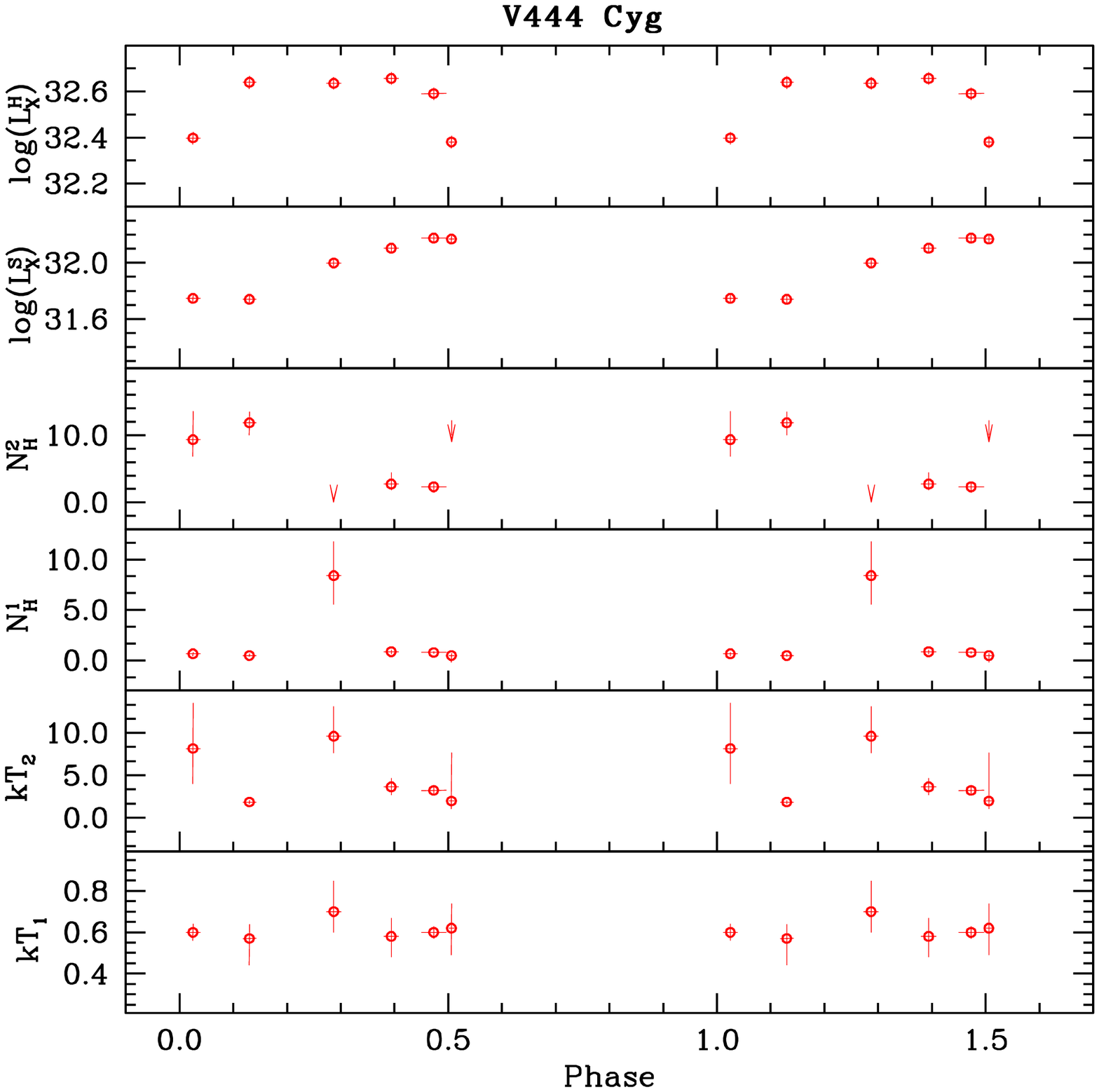}}
\subfigure[]{\includegraphics[width=3.0in]{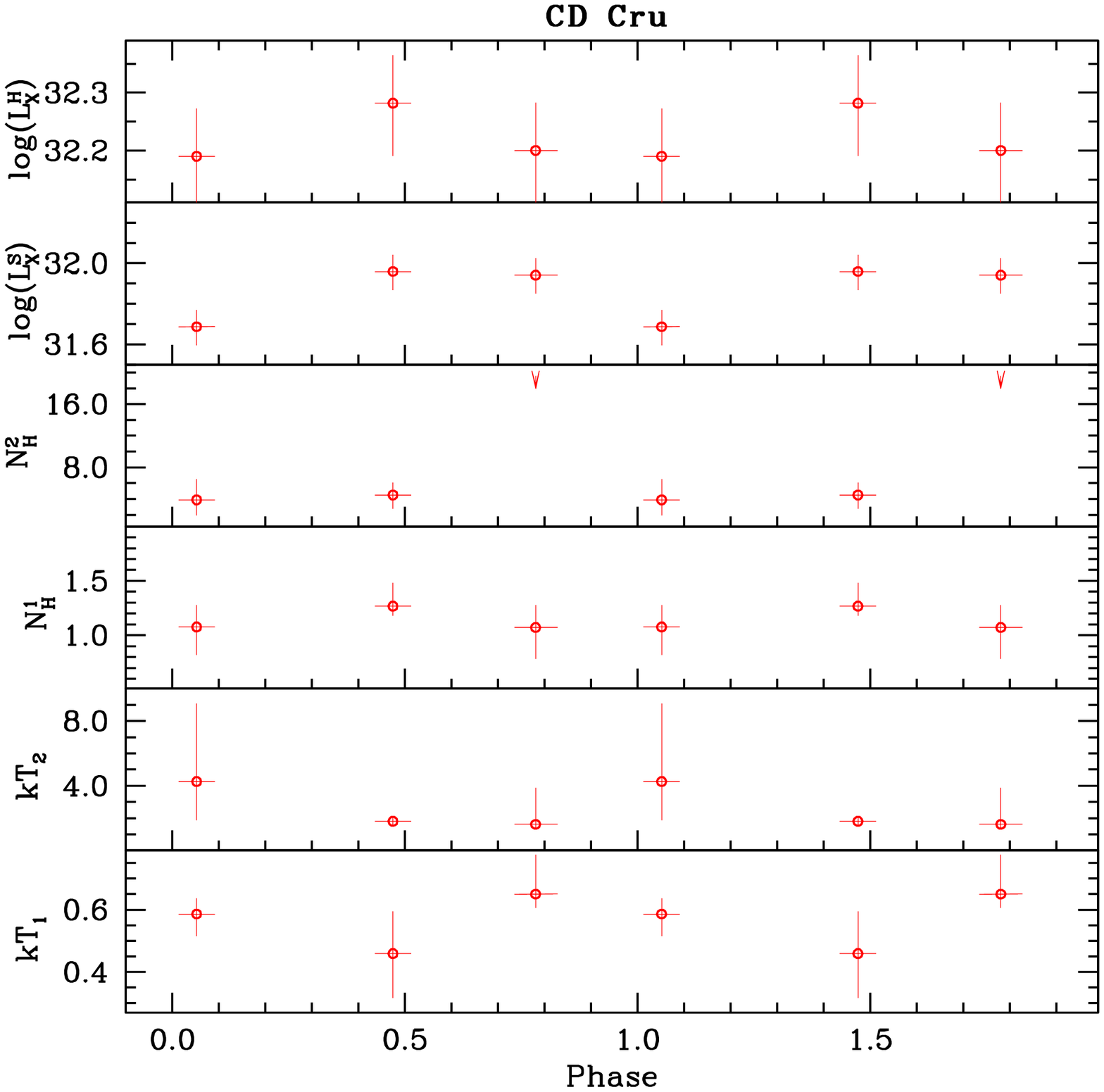}}
\caption{Variation of $\rm{L_X^S}$, $\rm{L_X^H}$, $\rm{N_H^1}$, $\rm{N_H^2}$, $\rm{kT_1}$ and $\rm{kT_2}$ as a function of orbital phase of the the stars (a) V444 Cyg and (b) CD Cru.}
\label{fig:ana_phase}
\end{figure}
\subsection {Evolution of spectral parameters} \label{sec:spt_evo}

The spectral analysis at different phases of the WR binaries provides the
dependence of the best fit values of parameters on orbital phase.
The variation of the soft ($\rm{L_X^S}$) and hard ($\rm{L_X^H}$) band X-ray luminosities, column densities corresponding to the cool ($\rm{N_H^1}$) and hot ($\rm{N_H^2}$) temperature components, and cool(kT$_1$) and hot (kT$_2$) temperatures as a function of orbital phases of V444 Cyg and CD Cru are shown in Fig. \ref{fig:ana_phase}(a) and \ref{fig:ana_phase}(b), respectively. In the case of V444 Cyg, the $\rm{L_X^H}$ was found to be minimum during the  primary  and secondary eclipses and maximum outside the eclipse. It was constant during the phase 0.13 to 0.47. However, the $\rm{L_X^S}$ was found to be minimum during the primary eclipse only. At the phase of 0.5 the  $\rm{L_X^S}$ was found to be maximum. Both  $\rm{L_X^H}$ and $\rm{L_X^S}$ were found to be minimum during the phase 0.0 and maximum at phase 0.5 in the WR binary CD Cru.
The cool component ($\rm{kT_1}$)
was found to be constant with a mean value of 0.61$\pm$0.05 keV and
0.57$\pm$0.10 keV for V444 Cyg and CD Cru, respectively.
The hard energy component ($\rm{kT_2}$) was varied from a minimum value of  1.88 keV  to a maximum value of  9.61 keV  for V444 Cyg. It was maximum at primary eclipse and at the phase of 0.29, and minimum during the secondary eclipse and at the phase of 0.13.
For CD Cru, $\rm{kT_2}$ varies from a minimum value 1.63  keV during outside the eclipse
to a maximum value of  4.27 keV at primary eclipse.
For the star V444 Cyg, $\rm{N_H^1}$ was maximum at phase 0.29,
otherwise it was  constant at all phases.
However, $\rm{N_H^1}$ was  found to be constant throughout the orbital phase with a mean value of 1.14$\pm$0.11 $\times \rm{10^{22}~cm^{-2}}$ for CD Cru.
For V444 Cyg, $\rm{N_H^2}$  was increased from phase 0.0 to its maximum value at phase 0.13.
Afterward it was decreased and became almost constant from phase 0.29  to phase 0.47.
For CD Cru, $\rm{N_H^2}$ was minimum at phase 0.0 and maximum at phase 0.78.

\begin{figure*}
\centering
\includegraphics[width=6.0in, height=5.0in]{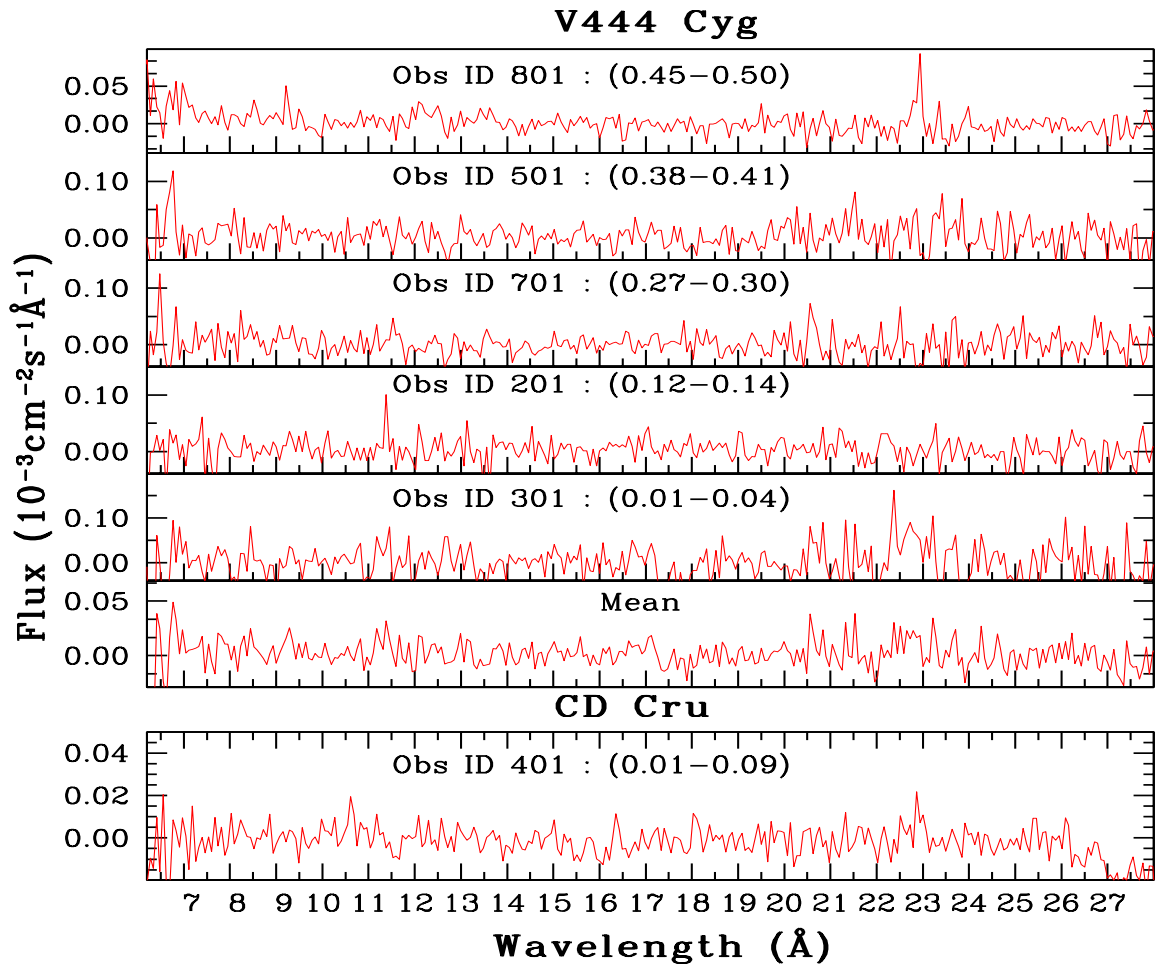}
\caption{RGS spectra of the WR binaries at different binary phase for the stars V444 Cyg (upper panel) and CD Cru (lower panel).}
\label{fig:rgs}
\end{figure*}
\subsection{RGS spectra} \label{sec:an_rgs}

We combined the first and second order spectra of detector RGS1 and RGS2 to inspect
the main spectral lines.
We obtained the RGS fluxed spectra using task {\sc rgsfluxer}
and is shown in Fig.~\ref{fig:rgs} for different phases for both the binaries.
We identified the prominent emission lines for V444 Cyg at wavelengths
Mg XII (6.50 \AA; 6.74 \AA), Mg XI (9.23 \AA), Fe XVIII (11.40 \AA) and
Ca XVI (22.61 \AA, 21.43 \AA, 22.30 \AA).
Individual lines show intensity variations from spectrum to spectrum.
Unfortunately, our observations are at the limit of detectability with the data available
with poor count statistics. Therefore, the signal-to-noise ratio of individual
lines are not sufficient to perform a detailed quantitative analysis of the
RGS line spectrum for any of the massive stars.

\section{Discussion}\label{sec:discuss}
We have analyzed the X-ray emission properties of two
WR binaries with strong stellar winds.
We discuss below the implications of the X-ray results.

\subsection{Phase-locked X-ray variability}\label{sec:discuss_lt}

The X-ray temporal and spectral analyses of WR binaries V444 Cyg and
CD Cru showed a phase-locked phenomenon.
The phase resolved X-ray spectroscopic observations show that the soft energy
component  and the corresponding $\rm{N_H^1}$ were also found to be nearly
 constant throughout the  binary phases of both stars.
The $\rm{L_X^S}$ was  found to be minimum during the  primary eclipse only. These results lead to the radiative wind shock origin of soft energy component.
In the case of CD Cru, the modulation of $\rm{L_X^S}$  well matches with the optical light curve reported by Moffat et al. (1990).  Lamontagne et al. (1996) have reported a similar shape of optical photometric light curves for 13 WR binaries including CD Cru and they said that it could be due to the "atmospheric eclipse".  
The X-ray flux in the hard energy band was found to be minimum at both primary and secondary eclipse for both stars.  This implies  that the hard energy component could be due to the presence of wind-wind collision zone which is located somewhere in between the O-type and WR star, and therefore, minimum when
either of the star in the binary system is eclipsed.
A similar behavior of X-ray light curves was seen in the EINSTEIN, ROSAT and ASCA observations of V444 Cyg (Moffat et al. 1982; Corcoran et al. 1996 ; Maeda et al. 1999).  The  hot temperature
($\rm{kT_2}$) and the corresponding $\rm{N_H^2}$ were also varied with orbital  phases of both binaries.  The eccentricity of both binary systems is almost zero,
therefore, the variation in the temperature corresponding to the hard energy component could be the result of variation in $\rm{N_H^2}$, i.e., varying optical depth (see \S\ref{sec:intro}).  This further supports the  wind-wind collision phenomenon in these binaries.

\subsection{X-ray Temperatures of Plasma}\label{sec:discuss_spt}

The X-ray spectra of WR binaries are well defined by two temperature plasmas.
The values of  $\rm{log(L_X/L_{bol})}$ in the total energy band  are found
to be -6.6 and -7.2  in for V444 Cyg and CD Cru, respectively,
which are similar to those for other WR (WN) stars observed
from {\sc XMM-Newton} and {\sc CHANDRA} (G$\ddot{u}$del \& Naz$\acute{e}$ 2009).
The temperature of the cool component of WR binaries
are comparable to other WR binaries e.g., 0.56--0.67 keV for WR1 (Ignace, Oskinova \& Brown 2003)
and  0.7--0.8 keV for WR 147 (Skinner et al. 2007). However,
the temperatures corresponding to hot energy component o
f these WR binaries are slightly more than some of the similar binaries.
The possible mechanisms for generation of X-rays are discussed below.

\subsubsection{ Instabilities driven radiative wind shocks }
The wind-shock model predicts the intrinsic instability of the line driving mechanism.
The standard model estimates the shock velocities
 by the relation  $\rm{kT_{sh}=1.95 \mu v^2_{shock}}$ (Lucy 1982; Luo, McCray \& Maclow 1990).
The temperatures of the cool energy components are found to be
almost similar ($\sim$ 0.6 keV) for both the binary systems (see Table~\ref{tab:V444Cyg_fit} and
Table~\ref{tab:CDCru_fit}).
Adopting the mean particle weight $\rm{\mu\approx}$1.16 (Skinner et al. 2007) for WN stars and
$\rm{\mu\approx}$0.62 for O-type stars (Cassinelli et al. 2008), we derived the
"average" value of shock velocities $\approx$ 515 $\rm{km~s^{-1}}$ for WN stars
and $\approx$ 704 $\rm{km~s^{-1}}$ for O-type stars corresponding to the cool component.
These values are about a factor of 2 larger than those predicted by radiative shock  model
 of Lucy (1982). However, the advance version of the wind shock model
by Owocki et al. (1988) predicts  X-ray emission up to 1 keV.
Moreover, the temperatures of the cool energy component
and $\rm{N_H^1}$ are not varying with orbital phase.
Therefore, it appears that the cool energy component from  V444 Cyg and CD Cru could be
generated by either of the binary components and may be explained
by instabilities in radiation-driven wind shocks.

\subsubsection{Magnetically Confined Wind Shock}
Babel \& Montmerle (1997) have suggested that the presence of magnetic fields confine the wind,
which may be an important ingredient for the production of X-ray emission (Babel \& Montmerle 1997).
The degree of confinement of wind by the magnetic field is
derived in terms of a confinement  parameter $\rm{ \Gamma=B^{2}_{0} R^{2}_{\ast}/\dot{M}v_{\infty} }$,
where  $\rm{B_0}$ is the surface equatorial magnetic field strength (ud-Doula \& Owocki 2002).
For a confined wind  model $\rm{\Gamma>>1}$.
Using the values of stellar radius ($\rm{R_{\ast}}$) and stellar mass loss rate ($\rm{\dot{M}v}$)
as given in  Table~\ref{tab:par_str} for V444 Cyg binary components,
 the minimum magnetic field required to confine the wind is derived to be 0.14 kG for the primary
and 1.25 kG  for the secondary.
Similarly, for CD Cru, we derived minimum magnetic field of 0.16 kG and 1.96 kG
for the primary  and the secondary components, respectively.
Based on the formula given by  Babel \& Montmerle  (1997) and using above estimated values of B, the X-ray luminosity are estimated to be $10^{34.66}$~\egs~and $10^{35.74}$ ~\egs~for  primary  and the secondary components of V444 Cyg, and
$10^{35.06}$~\egs~and  $10^{36.8}$ ~\egs~for primary  and the secondary components of CD Cru.
Such a high X-ray luminosity has not been observed from any of the WR binaries.
However, smaller fields can exist and the confinement may be limited to just above the magnetic equatorial plane.  

\subsubsection{Colliding wind shock model}
  \input{model.tab}
Assuming that shock velocities have reached to the terminal velocities,
the standard model predicts a maximum temperature
for the shocked region from winds of massive stars by
the relation  $\rm{kT^{max}_{sh}=1.95 \mu v^2_{\infty}}$ (Luo, McCray \& Maclow 1990).
The nominal wind parameters of massive stars (see Table~\ref{tab:par_str})
and the mean particle weight $\rm{\mu}$ for the WN stars and for the O-type stars give the maximum temperature
generated by the individual components of binaries.
The derived values are 7.80 keV + 7.21 keV for V444 Cyg (O6+WN5) and 10.88 keV + 13.69 keV for CD Cru (O5+WN6).
The maximum observed temperature at phase 0.29 corresponding to hot component is  found to be similar to that of maximum possible values in the case of V444 Cyg.
However, in CD Cru maximum observed temperature is lesser than that of maximum possible values.

The phase locked variability of  the hard energy component shows that the observed hot X-ray temperatures
in massive  binaries could originate from the collision of stellar winds.
The distances from the stars where these
winds meet are derived using the relations (De Becker 2007; Stevens, Blondin \& Pollock 1992),

\begin{equation}
\rm{  r_{OB}=\frac {1}  {1+\eta^{1/2}}  D}
\end{equation}

\noindent
where $\rm{r_{OB}}$ and D are the distance to the
collision zone from the primary and the separation between the binary components, respectively.
The wind momentum ratio $\rm{\eta}$ is expressed as,

\begin{equation}
\rm{\eta=\frac{\dot{M_2}v_{\infty,2}}{\dot{M_1}v_{\infty,1}}}
\end{equation}

\noindent
where $\rm{\dot{M_1}}$ and $\rm{\dot{M_2}}$
are mass loss rates of primary and secondary components, respectively, and
$\rm{v_{\infty,1}}$ and $\rm{v_{\infty,2}}$ are
terminal velocities of primary and secondary components, respectively.
Using the above relations, we derived wind momentum ratios of 7.03 and 24.60
for the stars V444 Cyg and CD Cru , respectively.
Distances to the
collision zone from the primary are estimated to be 0.27 and 0.17 times of their corresponding
binary separation (D) for V444 Cyg and CD Cru, respectively.
Therefore, the collision zone exists to be very close to the primary O-type star in case of
both the binaries. Further, the collision wind zone will produce a bow shock around the O-type star 
, the shock volume and the emission measure are dominated by the WR winds, therefore, 
X-ray emitting plasma also shows non-solar abundances.

\begin{figure}
\centering
\includegraphics[width=3.5in, height=3.5in]{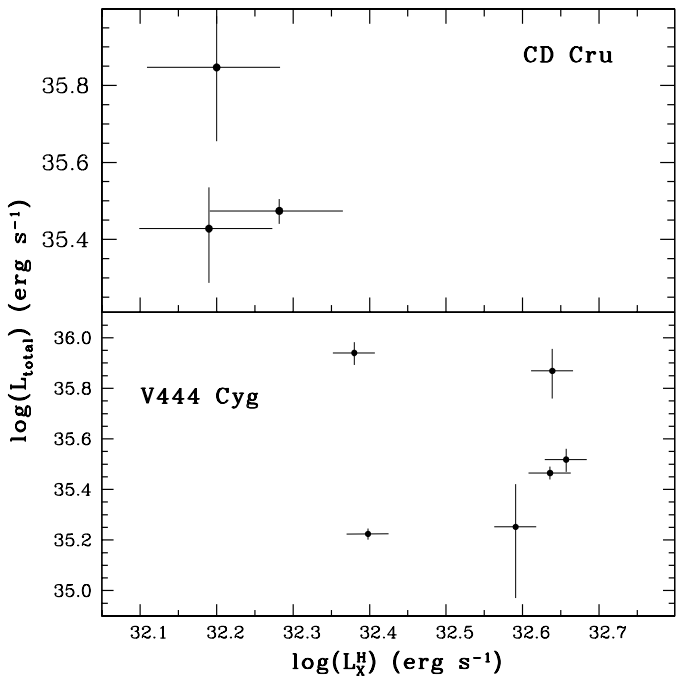}
\caption{Observed vs Model X-ray luminosity.}
\label{fig:model}
\end{figure}

 The gas in a colliding wind  region could be either adiabatic or radiative and depends upon
the value of cooling parameter ($\chi$), which is defined as the ratio of the cooling time ($\rm{t_{cool}}$) of the shocked gas 
to the escape time ($\rm{t_{esc}}$) from
the intershock region (Stevens, Blondin \& Pollock 1992).

\begin{equation}
\rm {\chi = \frac{t_{cool}}{t_{esc}} = \frac{v^4_3 d_7} {\dot{M}_{-7}}}
\end{equation}

\noindent
where 
$\rm{v_{3}}$ is the pre-shock velocity in the unit of 10$^3$ $\rm{km~s^{-1}}$,
$\rm{d_{7}}$ is the distance from stellar center to the shock in the unit of 10$^7$ km and
$\rm{\dot{M}_{-7}}$ is the mass loss rate in the unit of $\rm{10^{-7}M_{\odot} yr^{-1}}$.
Using the parameters given in Table~\ref{tab:par_str} and the values from above estimation, we derived cooling parameters
$\rm{\chi_1}$ and $\rm{\chi_2}$ for primary and secondary components, respectively.
The estimated values of $\rm{\chi_1}$ and  $\rm{\chi_2}$ are given Table~\ref{tab:model}.
The winds from WR stars are found to be clearly radiative ($\rm{\chi<1}$) for both the binaries.
The intrinsic luminosity can be estimated for each component of the binary using the following relation
(Pittard  \& Stevens 2002) :

\begin{equation}
\rm {L_X = 0.5~\Xi~\dot{M}~v^2 }
\end{equation}

\noindent
where $\rm{\Xi}$ accounts for a geometrical and inefficiency factors and
v is the wind speed at contact surface (i.e. pre-shock velocities derived from observed temperatures).
The values of $\rm{\Xi_{1}}$ (0.403 for V444 Cyg ; 0.564 for CD Cru)
and $\rm{\Xi_{2}}$ (0.033 for V444 Cyg ; 0.0042 for CD Cru) for primary
and secondary stars, respectively, are taken from
Pittard  \& Stevens (2002).
The estimated values of intrinsic luminosities of each component and the total intrinsic
luminosities of binary systems are given in Table~\ref{tab:model}.
These values are nearly 3 orders of magnitude higher than those
of the  observed values. A similar results have been found for other WR binaries, e.g.,  HD 159176 (De Becker et al. 2004); WR 147 (Skinner et al. 2007).
De Becker et al. (2004) suggested that the disagreement between observed and theoretical predictions could possibly be explained by (a) the kinetic power of the collision should be considered as an upper limit on the X-ray luminosity, (b) higher value of the parameter $\rm{\eta}$ should be considered, (c) diffusive mixing between hot and cool material is likely to exist due to the instability of the shock front, and (d) orbital effects should also be included to study such systems.    
The observed and theoretically predicted values of X-ray luminosities are
shown in Fig.~\ref{fig:model}. This model also predicts the phase-locked variations
 which can be seen in Fig.~\ref{fig:ana_phase}.
Therefore, it appears that the hard energy component is most likely associated with wind collision zone. 

\subsubsection{ Non-thermal emission}

There is no strong justification for invoking non-thermal origin of X-ray emission since
the observed X-ray emission at higher energies can be fitted satisfactorily with a
thermal model (see Fig.~\ref{fig:spt_V444Cyg} and Fig.~\ref{fig:spt_CDCru}).
However, the present data is limited up to 10 keV and it is difficult to eliminate
the contribution of non-thermal components from the spectra which is dominated by
the presence of the strong thermal emission, and the non-thermal emission is
overwhelmed by the thermal emission (De Becker 2007).
Therefore, we can state that the spectra of massive stars in the present sample are fairly well explained by
two-temperature plasma models.

\section{Summary}\label{sec:con}
We have analysed X-ray temporal and spectral properties of two WR binaries  V444 Cyg and CD Cru observed with high sensitivity EPIC instruments on board the XMM-Newton satellite.
The X-ray light curves in the soft and the hard energy band show the phase locked variability. Both primary and secondary minima were seen in the  hard band X-ray light curves of both binaries. However, in the soft X-ray light curve only primary minimum was seen. This implies that the hard energy component could be originating from the wind-wind collision zone.
The X-ray spectra of both  WR binaries show strong
absorption below $\approx$ 1 keV and  a clear evidence of high temperature
plasma, manifested by a visible Fe $\rm {K\alpha}$ emission-line complex.
The study of X-ray spectra reveals cool as well as hot temperature plasma components
of binary stars which are fitted consistently with two-temperature plasma models.
The cooler plasma component was found to be constant at all phases with mean value of $\sim$0.6 keV  for both binaries. The presence of cooler component  could be due to the distribution of small-scale shocks in the radiation-driven
outflows from either the primary or the secondary star in the binary systems.
  The temperature of hot plasma component and the corresponding column density  were found to be variable during the orbital cycle of both binaries. The variation in temperature of hot plasma could be due to the varying circumstellar optical depth along the line of sight towards the shock as star revolves around each other. The maximum value of hot plasma was found to be  lower than the hottest plasma possible in the binary systems as predicted by colliding wind theory for  short-period binaries.  However, the predicted values of X-ray luminosities are $\sim 3$ orders of magnitude more than those
of the  observed values and can not be accounted in terms of observational errors.


\section*{acknowledgments}

We thank referee Dr. Marc Gagn$\acute{e}$ for his useful suggestions. 
We would like to thank Dr. Maheshwar Gopinathan for the critical reading of the manuscript. 
 This publication makes use of data products from XMM-Newton archives using the high energy astrophysics science archive research center which is established at Goddard by NASA. We acknowledge XMM-Newton Help Desk for their remarkable support in X-ray data analysis.  This research has also made use of data Simbad and VizieR catalogue access tool, CDS, Strasbourg, France.

\input{ref.tex}

\end{document}

%% file: stellar_par.tab
\begin{table*}
\caption{Stellar parameters of WR binaries V444 Cyg and CD Cru.}
\label{tab:par_str}      
\begin{tabular}{lcccc}       
\hline
Parameters                                      & \multicolumn{2}{c}{V444 Cyg} & \multicolumn{2}{c}{CD Cru}   \\
                                                &                    & Ref$^a$ &                 & Ref$^a$     \\
\hline
Other Names                                     & HD 193576; WR 139  &          & HD 311884 ; WR 47             \\
Spectral Type                                   & O6 III-V + WN5     & 1        & O5 V + WN6      & 1             \\
Mass ($\rm{M_{\odot}}$)                         & 25 + 10            & 2        & 57 + 48         & 10             \\
V (mag)                                         & 7.94               & 3        &  10.89          & 11             \\
Distance   (pc)                                 & 1900               & 3        &  3000           & 12            \\
$\rm{A_V}$ (mag)                                & 2.48               & 3        &  3.56           & 12            \\
Orbital Period  (days)                          & 4.21               & 2        &  6.24           & 13           \\
Binary   separation   ($\rm{R_{\odot}}$)        & 38                 & 2        &  68             & 10           \\
inclination (\deg)                              & 78                 & 4        &  70             & 10              \\
eccentricity                                    & 0.03               & 5        &  0              & 14          \\
$\rm{\dot{M}}$ ($\rm{10^{-5}~M_\odot~yr^{-1}}$) & 0.06 + 0.60        & 6,4      & 0.10 + 3.00     & 15,10       \\
$\rm{v_{\infty}}$ ($\rm{km~s^{-1}}$)            & 2540 + 1785        & 7,8      & 3000 + 2460     & 15,8       \\
$\rm{log(L_{bol})~erg~s^{-1}}$                  & 39.24 + 38.56      & 9        & 39.53  + 38.69  & 9           \\
Radius ($\rm{R_{\odot}}$)                       & 10 + 3             & 5        & 57 + 48         & 10           \\
Cluster Membership                              & Berkeley 86        & 3        & Hogg 15         & 11           \\
\hline
\end{tabular}
\newline
\noindent
$^a$ References:
1.  van der Hucht et al. (1981)
2.  M$\ddot{u}$nch (1950)
3.  Massey, Johnson \& De Gioio-Eastwood (1995)
4.  Kurosawa, Hillier \& Pittard (2002)
5.  Cherepashchuk, Eaton \& Khaliullin (1984)
6.  Marchenko et al. (1994)
7.  Underhill \& Fahey (1987)
8.  Nugis \& Lamers (2000)
9. Landolt-B$\ddot{o}$rnstein (1982)
10. Moffat et al. (1990)
11. Moffat (1974)
12. Sagar, Munari \& Boer (2001)
13. Niemela, Massey, Conti (1980)
14. Cherepashchuk \& Karetnikov (2003)
15. Repolust, Puls \& Herrero (2004) (typical value for the spectral-type)
\end{table*}

%% file: xray_observations.tab
\begin{table*}
\scriptsize
\caption{Journal of XMM-Newton Observations of the objects in the sample.}             
\label{tab:xray_observations}      
\begin{tabular}{lllllll}       
\hline 
Object Name  & Observation        & Exposure           &~~~~~~~~~ Start time  & EPIC filter  & Offset from      & Orbital  \\
              &     ID            &  Time  (sec)       & ~~~~~~~~UT(hh:mm:ss) &              & Target (arcmin)  & Phase    \\
\hline
V444 Cyg            &    0206240201   &   10032  &19 May 2004 12:35:42   & Thick    & 0.001    &   0.13 (0.12--0.14)\\
                    &    0206240301   &   11914  &27 May 2004 12:01:39   & Thick    & 1.120    &   0.03 (0.01--0.04)\\
                    &    0206240401   &   11915  &29 May 2004 12:01:49   & Thick    & 1.162    &   0.51 (0.50--0.51)\\
                    &    0206240501   &   11918  &06 June 2004 11:36:30  & Thick    & 1.111    &   0.39 (0.38--0.41)\\
                    &    0206240701   &   11911  &14 June 2004 10:58:29  & Thick    & 1.133    &   0.29 (0.27--0.30)\\
                    &    0206240801   &   19921  &27 Oct  2004 23:47:44  & Thick    & 1.093    &   0.47 (0.45--0.50)\\
\hline
CD Cru              &    0109480101   &   53046  &03 July 2002 15:51:56  &  Thick   & 2.812    &   0.78 (0.73--0.83)  \\ 
                    &    0109480201   &   48015  &26 Aug  2002 21:55:14  &  Thick   & 3.848    &   0.47 (0.43--0.51) \\ 
                    &    0109480401   &   53040  &21 Jan  2003 01:07:38  &  Thick   & 4.969    &   0.05 (0.01--0.09) \\ 
\hline
\end{tabular}
\end{table*}

%% file: bestfit_WR_V444Cyg.tab
\begin{landscape}
\begin{table}
\tiny
\caption {Spectral parameters derived from the best fit {\sc vpshock} model and two-temperature {\sc vapec} plasma models for V444 Cyg.}
\label{tab:V444Cyg_fit}      
\begin{tabular}{lllllllllllll}       
\hline

\multicolumn{13}{c}{V444 Cyg} \\
\hline                                                                             
Obs ID                              & \multicolumn{2}{c}{0206240201}                     & \multicolumn{2}{c}{0206240301}               & \multicolumn{2}{c}{0206240401}                                              & \multicolumn{2}{c}{0206240501}                                                 &  \multicolumn{2}{c}{0206240701}                                             &  \multicolumn{2}{c}{0206240801} \\
Phase                               & \multicolumn{2}{c}{(0.12--0.14)}                   &\multicolumn{2}{c}{(0.01--0.04)}              & \multicolumn{2}{c}{(0.50--0.51)}                                            & \multicolumn{2}{c}{(0.38--0.41)}                                               & \multicolumn{2}{c}{(0.27--0.30)}                                            & \multicolumn{2}{c}{(0.45--0.50)}\\
\hline                                                                             
Model                               & vpshock  &    2Tvapec                              & vpshock  &    2Tvapec                        &     vpshock                               &    2Tvapec                      &   vpshock                                 &   2Tvapec                          &   vpshock                                &    2Tvapec                       &   vpshock                                &    2Tvapec         \\               
\hline                                                                             
 $\rm{N_{H}^{1}}$                   & 0.75      & $\rm{0.48^{+0.20}_{-0.12}}$            &  0.280   & $\rm{0.64^{+0.16}_{-0.24}}$       & $\rm{0.17^{+1.46}_{-0.18}}$               & $\rm{0.47^{+0.26}_{-0.68}}$     & $\rm{1.46^{+0.30}_{-0.18}}$               & $\rm{0.85^{+0.20}_{-0.25}}$        &  $\rm{1.71^{+0.15}_{-0.40}}$             & $\rm{8.40^{+3.42}_{-2.88}}$      &  $\rm{0.43^{+0.15}_{-0.19}}$             & $\rm{0.78^{+0.12}_{-0.11}}$         \\            
 $\rm{N_{H}^{2}}$                   &           & $\rm{11.86^{+1.66}_{-1.88}}$           &          & $\rm{9.35^{+4.23}_{-2.50}}$       &                                           & $<$12.20                        &                                           & $\rm{2.74^{+1.71}_{-0.91}}$        &                                          & $\rm{<0.17}$                     &                                          & $\rm{2.33^{+0.76}_{-0.84}}$         \\           
 $\rm{kT_1}$                        & 79.80     & $\rm{0.57^{+0.07}_{-0.13}}$            &  79.90   & $\rm{0.60^{+0.04}_{-0.04}}$       & $\rm{2.33^{+0.74}_{-0.48}}$               & $\rm{0.62^{+0.12}_{-0.13}}$     & $\rm{3.14^{+0.92}_{-0.43}}$               & $\rm{0.58^{+0.09}_{-0.10}}$        &  $\rm{3.48^{+0.55}_{-0.46}}$             & $\rm{0.70^{+0.15}_{-0.10}}$      &  $\rm{4.21^{+0.70}_{-0.78}}$             & $\rm{0.60^{+0.03}_{-0.03}}$         \\            
 $\rm{kT_2}$                        &           & $\rm{1.85^{+0.22}_{-0.26}}$            &          & $\rm{8.16^{+5.35}_{-4.16}}$       &                                           & $\rm{1.97^{+5.70}_{-0.95}}$     &                                           & $\rm{3.64^{+1.01}_{-0.97}}$        &                                          & $\rm{9.61^{+3.51}_{-1.97}}$      &                                          & $\rm{3.22^{+0.68}_{-0.34}}$         \\            
 $\rm{\tau }$                       & 0.23      &                                        &  0.18    &                                   & $\rm{0.38^{+0.82}_{-0.19}}$               &                                 & $\rm{0.73^{+0.23}_{-0.20}}$               &                                    &  $\rm{0.58^{+0.18}_{-0.14}}$             &                                  &  $\rm{0.35^{+0.10}_{-0.07}}$             &                                     \\            
 $\rm{EM_1 }$                       & 0.19      & $\rm{<5.32}$                           &  0.17    & $\rm{<22.54}$                     & $\rm{<4.25}$                              & $<$288.07                       & $\rm{1.72^{+0.65}_{-0.80}}$               &  $\rm{29.33^{+60.06}_{-7.17}}$     &  $\rm{<35.92}$                           & $3.51^{+20.73}_{-2.52}$          &  $\rm{<11.77}$                           & $\rm{30.32^{+50.70}_{-22.53}}$        \\            
 $\rm{EM_2 }$                       &           & $5.10^{+0.72}_{-0.81}$                 &          & $13.60^{+122.46}_{-6.11}$          &                                          & $<$36.44                        &                                           &  $\rm{<74.03}$                     &                                          & $<25.367$                         &                                          & $\rm{16.28^{+48.01}_{-12.86}}$        \\            
 He                                 &  763.00   & $\rm{<30.15}$                          & 518.90   & $\rm{<0.14}$                      &  $\rm{<9997.63}$                          & $\rm{\approx 0.00}$             &  $\rm{<951.62}$                           & $\rm{<728.99}$                     &   $\rm{617.61^{+9380.65}_{-201.57}}$     & $\rm{>40.18}$                    &   $\rm{9973.5^{+9520.95}_{-199.229}}$    & $\rm{>20.04}$                      \\            
 C                                  & 9998.91   & $\rm{<693.28}$                         &   0.00   & $\rm{\approx 0.00}$               &  $\rm{\approx 0.00}$                      & $\rm{\approx 0.00}$             &  $\rm{\approx 0.00}$                      & $\rm{<258.19}$                     &   $\rm{\approx 0.00}$                    & $\rm{\approx 0.00}$              &   $\rm{<9999.97}$                        & $\rm{<109.71}$                      \\            
 N                                  &  650.45   & $\rm{<651.73}$                         &   0.00   & $\rm{ \approx 0.00}$              &  $\rm{<9998.63}$                          & $\rm{\approx 0.00}$             &  $\rm{\approx 0.00}$                      & $\rm{\approx 0.00}$                &   $\rm{>400.059}$                        & $\rm{\approx 0.00}$              &   $\rm{<98.928}$                         & $\rm{<885.30}$                      \\            
 O                                  &  191.35   & $\rm{377.38^{+50.80}_{-33.99}}$        &   2.13   & $\rm{<66.17}$                     &  $\rm{<25.84}$                            & $\rm{\approx 0.00}$             &  $\rm{1114.90^{+9667.15}_{-201.29}}$      & $\rm{<191.51}$                     &   $\rm{3054.69^{+7682.71}_{-48.20}}$     & $\rm{<316.85}$                   &   $\rm{<37.34}$                          & $\rm{<18.09}$                       \\            
 Ne                                 &   68.34   & $\rm{<80.79}$                          &  33.83   & $\rm{<20.27}$                     &  $\rm{<11.42}$                            & $\rm{<150.67}$                  &  $\rm{<58.76}$                            & $\rm{10.13^{+64.95}_{-3.96}}$      &   $\rm{<61.40}$                          & $\rm{<446.78}$                   &   $\rm{<2.96}$                           & $\rm{<0.49}$                        \\            
 Mg                                 &   0.00    & $\rm{31.77^{+130.88}_{-26.32}}$        &  76.34   & $\rm{38.42^{+365.20}_{-7.61}}$    &  $\rm{40.46^{+289.49}_{-17.30}}$          & $\rm{29.21^{+225.47}_{-10.36}}$ &  $\rm{5.34^{+9.53}_{-4.66}}$              & $\rm{2.83^{+15.40}_{-2.13}}$       &   $\rm{<46.27}$                          & $\rm{<658.42}$                   &  $\rm{13.17^{+6.66}_{-4.49}}$            & $\rm{1.43^{+2.98}_{-0.62}}$        \\            
 Si                                 &   0.00    & $\rm{62.63^{+152.45}_{-40.66}}$        & 142.88   & $\rm{37.36^{+344.36}_{-9.20}}$    &  $\rm{78.56^{+454.70}_{-35.98}}$          & $\rm{97.03^{+220.92}_{-13.96}}$ &  $\rm{21.20^{+56.08}_{-3.95}}$            & $\rm{3.19^{+80.86}_{-5.29}}$       &   $\rm{73.63^{+37.30}_{-24.93}}$         & $\rm{120.13^{+411.08}_{-68.09}}$ &  $\rm{47.98^{+16.25}_{-9.83}}$           & $\rm{4.59^{+33.75}_{-1.00}}$       \\            
 S                                  &  999.97   & $\rm{38.33^{+135.50}_{-17.19}}$        & 199.06   & $\rm{73.74^{+664.80}_{-19.03}}$   &  $\rm{217.19^{+460.11}_{-133.78}}$        & $\rm{58.63^{+491.05}_{-19.90}}$ &  $\rm{25.91^{+85.94}_{-8.20}}$            & $\rm{<72.61}$                      &   $\rm{212.54^{+109.78}_{-33.55}}$       & $\rm{203.53^{+250.52}_{-91.46}}$ &  $\rm{184.70^{+855.09}_{-40.00}}$        & $\rm{44.18^{+38.89}_{-3.08}}$      \\            
 Ar                                 &    6.02   & $\rm{<119.86}$                         &   6.02   & $\rm{<64.00}$                     &  $\rm{\approx 0.00}$                      & $\rm{>2.77}$                    &  $\rm{\approx 0.00}$                      & $\rm{28.83^{+32.02}_{-16.22}}$     &   $\rm{\approx 0.00}$                    & $\rm{246.82^{+380.31}_{-194.31}}$&  $\rm{\approx 0.0}$                      & $\rm{23.98^{+67.11}_{-3.75}}$      \\            
 Ca                                 & 9991.50   & $\rm{<11.46}$                          &2994.71   & $\rm{<97.37}$                     &  $\rm{<995.93}$                           & $\rm{<337.62}$                  &  $\rm{89.45^{+125.95}_{-70.88}}$          & $\rm{<37.60}$                      &   $\rm{631.16^{+9999.82}_{-260.60}}$     & $\rm{>182.96}$                   &  $\rm{<9997.92}$                         & $\rm{4.89^{+35.39}_{-4.28}}$       \\            
 Fe                                 &   92.50   & $\rm{20.95^{+101.17}_{-24.64}}$        & 196.91   & $\rm{66.99^{+620.40}_{-7.20}}$    &  $\rm{14.27^{+113.12}_{-9.40}}$           & $\rm{9.22^{+157.32}_{-8.12}}$   &  $\rm{23.31^{+105.67}_{-13.01}}$          & $\rm{1.93^{+12.98}_{-1.45}}$       &   $\rm{162.51^{+9836.23}_{-106.27}}$     & $\rm{142.33^{+124.11}_{-57.97}}$ &  $\rm{11.35^{+7.76}_{-7.19}}$            & $\rm{1.11^{+4.37}_{-0.31}}$         \\            
 $\rm{\chi^{2}_{\nu}(d.o.f.)}$      & 2.77(130) &  1.07 (127)                            & 2.38(79) &  1.28 (77)                        &  1.10 (27)                                &  0.77 (25)                      &  1.47 (200)                               &  1.38 (198)                        &   1.13 (168)                             &  1.08 (166)                      &    1.12(277)                             &  0.98 (275)                         \\            
 $\rm{L_X^T }$                      &           &  32.691                                &          &  32.485                           &                                           &  32.589                         &                                           &  32.764                            &                                          &  32.726                          &                                          &  32.732                              \\            
 $\rm{L_X^S}$                       &           &  31.739                                &          &  31.746                           &                                           &  32.170                         &                                           &  32.104                            &                                          &  31.999                          &                                          &  32.177                                   \\            
 $\rm{L_X^H}$                       &           &  32.639                                &          &  32.398                           &                                           &  32.380                         &                                           &  32.657                            &                                          &  32.636                          &                                          &  32.591                                   \\            
\hline
\end{tabular}
\newline
\noindent
{\it Note:} Here $\chi^2_\nu = \chi^2/\nu$, where $\nu$ is degrees of freedom. Errors are with 90\% confidence based on $\chi_{min}^2 + 2.71$.\\
\noindent
$\rm{N_{H}^{1}}$ and $\rm{N_{H}^{2}}$ are in the units of $\rm{10^{22}~cm^{-2}}$.\\
$\rm{kT_1}$ and $\rm{kT_2}$ are in the units of keV.\\
$\rm{\tau}$ is in the units of  $\rm{10^{11}~s~cm^{-3}}$.\\
$\rm{EM_1}$ and  $\rm{EM_2}$ are in the units of $\rm{10^{54}~cm^{-3}}$.\\
$\rm{L_X^T}$ luminosity in band 0.3--7.5 keV, $\rm{L_X^S}$ luminosity in band 0.3--2.0 keV  
and $\rm{L_X^H}$ luminosity in band 2.0--7.5 keV  are in the units of $\rm{erg s^{-1}}$.\\ 
\end{table}
\end{landscape}

%% file: bestfit_WR_CDCru.tab
\begin{table*} 
\scriptsize
\caption {Spectral parameters derived from the best fit {\sc vpshock} model and two-temperature {\sc vapec} plasma models for CD Cru. Units of the parameters are same as given in Table~\ref{tab:V444Cyg_fit}.}
\label{tab:CDCru_fit}      
\begin{tabular}{lllllll}       
\hline

\multicolumn{7}{c}{CD Cru} \\
\hline                                                                             
Obs ID                              & \multicolumn{2}{c}{0109480101}                     & \multicolumn{2}{c}{0109480201}               & \multicolumn{2}{c}{0109480401}   \\                                           
Phase                               & \multicolumn{2}{c}{(0.73--0.83)}                   &\multicolumn{2}{c}{(0.43--0.51)}              & \multicolumn{2}{c}{(0.01--0.09)} \\                                           
\hline                                                                             
Model                               & vpshock                              &    2Tvapec                             & vpshock                                      &    2Tvapec                        & vpshock     &    2Tvapec   \\                        
\hline                                                                             
 $\rm{N_{H}^{1}}$                   & $\rm{1.44^{+0.14}_{-0.22}}$          & $\rm{1.07^{+0.20}_{-0.29}}$            & $\rm{0.27^{+0.12}_{-0.10}}$                  & $\rm{1.27^{+0.21}_{-0.09}}$       & $\rm{0.78^{+0.19}_{-0.16}}$                             & $\rm{1.08^{+0.20}_{-0.26}}$          \\            
 $\rm{N_{H}^{2}}$                   &                                      & $\rm{<19.57}$                          &                                              & $\rm{4.49^{+1.57}_{-1.75}}$       &                                                         & $\rm{3.88^{+2.64}_{-1.96}}$           \\           
 $\rm{kT_1}$                        & $\rm{2.67^{+0.82}_{-0.82}}$          & $\rm{0.65^{+0.13}_{-0.04}}$            & $\rm{3.02^{+0.51}_{-0.42}}$                  & $\rm{0.46^{+0.14}_{-0.14}}$       & $\rm{7.69^{+7.08}_{-3.05}}$                             & $\rm{0.59^{+0.05}_{-0.07}}$          \\            
 $\rm{kT_2}$                        &                                      & $\rm{1.63^{+2.26}_{-0.23}}$            &                                              & $\rm{1.81^{+0.31}_{-0.29}}$       &                                                         & $\rm{4.27^{+4.81}_{-2.39}}$          \\            
 $\rm{\tau }$                       & $\rm{0.60^{+0.12}_{-0.13}}$          &                                        & $\rm{1.19^{+0.51}_{-0.35}}$                  &                                   & $\rm{0.25^{+0.07}_{-0.05}}$                             &                                      \\            
 $\rm{EM_1 }$                       & $\rm{>8.80\times10^4}$               & $\rm{0.55^{+25.35}_{-0.17}}$           & $\rm{0.03^{+0.62}_{-0.01}}$                  & $\rm{6.33^{+57.90}_{-0.30}}$      & $\rm{<0.44}$                                            & $<$270.41                              \\            
 $\rm{EM_2 }$                       &                                      & $\rm{<45.06}$                          &                                              & $\rm{0.82^{+0.49}_{-0.44}}$       &                                                         & $<$45.68                               \\            
 He                                 &  $\rm{493.48^{+9492.43}_{-421.12}}$  & $\rm{<202.72}$                         &  $\rm{5319.53^{+7609.94}_{-29.35}}$          & $\rm{<171.03}$                    &  $\rm{375.17^{+9580.50}_{-145.95}}$                     & $\rm{\approx 0.00}$                 \\            
 C                                  &  $\rm{<9999.98}$                     & $\rm{\approx 0.00}$                    &  $\rm{\approx 0.00}$                         & $\rm{<445.56}$                    &  $\rm{\approx 0.00}$                                    & $\rm{\approx 0.00}$                  \\            
 N                                  &  $\rm{\approx 0.0}$                  & $\rm{\approx 0.00}$                    &  $\rm{< 9999.53}$                            & $\rm{<805.06}$                    &  $\rm{\approx 0.00}$                                    & $\rm{\approx 0.00}$                  \\            
 O                                  &  $\rm{4947.23^{+7632.81}_{-5.07}}$   & $\rm{>137.01}$                         &  $\rm{< 100.83}$                             & $\rm{>137.33}$                    &  $\rm{< 421.57}$                                        & $\rm{\approx 0.00}$                  \\            
 Ne                                 &  $\rm{<195.20}$                      & $\rm{217.83^{+122.88}_{-107.12}}$      &  $\rm{< 200.47}$                             & $\rm{82.73^{+103.68}_{-65.56}}$   &  $\rm{< 45.27}$                                         & $\rm{<3.95}$                       \\            
 Mg                                 &  $\rm{110.50^{+52.64}_{-26.30}}$     & $\rm{84.33^{+33.16}_{-37.95}}$         &  $\rm{312.97^{+133.48}_{-105.30}}$           & $\rm{79.94^{+58.32}_{-32.51}}$    &  $\rm{64.24^{+115.36}_{-20.92}}$                        & $\rm{27.17^{+29.04}_{-0.64}}$     \\            
 Si                                 &  $\rm{256.53^{+52.64}_{-52.64}}$     & $\rm{101.23^{+54.24}_{-31.93}}$        &  $\rm{801.39^{+8604.51}_{-263.32}}$          & $\rm{163.73^{+92.97}_{-75.87}}$   &  $\rm{80.84^{+89.41}_{-33.29}}$                         & $\rm{6.46^{+29.71}_{-0.27}}$     \\            
 S                                  &  $\rm{298.46^{+105.28}_{-52.64}}$    & $\rm{191.89^{+81.16}_{-69.29}}$        &  $\rm{1090.01^{+438.26}_{-172.37}}$          & $\rm{125.29^{+81.78}_{-40.03}}$   &  $\rm{193.76^{+277.87}_{-84.88}}$                       & $\rm{21.85^{+207.92}_{-1.64}}$     \\            
 Ar                                 &  $\rm{ \approx 0.0}$                 & $\rm{<73.88}$                          &  $\rm{ \approx 0.00}$                        & $\rm{101.34^{+120.36}_{-73.27}}$  &  $\rm{ \approx 0.00}$                                   & $\rm{>1.32}$                        \\            
 Ca                                 &  $\rm{>862.33}$                      & $\rm{<318.32}$                         &  $\rm{<3889.83}$                             & $\rm{<262.68}$                    &  $\rm{<948.97}$                                         & $\rm{<5.50}$                      \\            
 Fe                                 &  $\rm{780.96^{+210.56}_{-421.12}}$   & $\rm{90.78^{+81.92}_{-78.33}}$         &  $\rm{<237.873}$                             & $\rm{86.23^{+79.64}_{-38.00}}$    &  $\rm{104.73^{+202.96}_{-58.14}}$                       & $\rm{5.38^{+50.63}_{-1.07}}$        \\            
 $\rm{\chi^{2}_{\nu}(d.o.f.)}$      &  1.06(164)                           &  1.01 (162)                            &  1.44(173)                                   &  1.32 (171)                       &  0.98(89)                                               &  0.95 (87)                           \\            
 $\rm{F_X}$                         &  0.23                                &  0.23                                  &  0.26                                        &  0.26                             &  0.19                                                   &  0.19                              \\            
 $\rm{L_X^T }$                      &                                      & 32.391                                 &                                              &  32.451                           &                                                         &  32.309                       \\            
 $\rm{L_X^S}$                       &                                      & 31.941                                 &                                              &  31.959                           &                                                         &  31.687                   \\            
 $\rm{L_X^H}$                       &                                      & 32.200                                 &                                              &  32.282                           &                                                         &  32.190                   \\            
\hline
\end{tabular}
\newline
\noindent
{\it Note:} Here $\chi^2_\nu = \chi^2/\nu$, where $\nu$ is degrees of freedom. Errors are with 90\% confidence based on $\chi_{min}^2 + 2.71$.
\end{table*}

%% file: model.tab
\begin{table*}
\caption{Derived values of wind parameters and intrinsic luminosities for binary systems.
The pre-shock velocities for each components are estimated using derived values of post-shock temperatures.
}
\label{tab:model}      
\begin{tabular}{cccccccc}       
\hline 
Object  &                  & $\rm{v_{pre-shock}}$                                  &  $\rm{\chi_1}$  & $\rm{\chi_2}$ & log( $\rm{L_{X1}}$)  & log($\rm{L_{X2}}$)  & log($\rm{L_{total}}$)  \\
Name    &   Phase          &     $\rm{km~s^{-1}}$                                  &                 &               &  $\rm{erg~s^{-1}}$   & $\rm{erg~s^{-1}}$   & $\rm{erg~s^{-1}}$      \\
\hline 

V444 Cyg&  0.01--0.04      &  $\rm{2598^{+745}_{-779}  +  1899^{+545}_{-570} }$    &    5.496        &  0.416        &    $\rm{35.711^{+0.109}_{-0.155}}$ &   $\rm{35.352^{+0.109}_{-0.155}}$ &     $\rm{35.869^{+0.087}_{-0.109}}$  \\
        &  0.12--0.14      &  $\rm{1237^{+71}_{-90}    +   904^{+52}_{-66}   }$    &    0.282        &  0.021        &    $\rm{35.067^{+0.024}_{-0.033}}$ &   $\rm{34.708^{+0.024}_{-0.033}}$ &     $\rm{35.224^{+0.021}_{-0.022}}$  \\
        &  0.27--0.30      &  $\rm{2819^{+475}_{-306}  +  2061^{+347}_{-223} }$    &    7.623        &  0.577        &    $\rm{35.782^{+0.068}_{-0.050}}$ &   $\rm{35.423^{+0.068}_{-0.050}}$ &     $\rm{35.940^{+0.043}_{-0.048}}$   \\
        &  0.38--0.41      &  $\rm{1735^{+226}_{-249}  +  1269^{+165}_{-182} }$    &    1.094        &  0.083        &    $\rm{35.361^{+0.053}_{-0.067}}$ &   $\rm{35.002^{+0.053}_{-0.067}}$ &     $\rm{35.518^{+0.043}_{-0.048}}$   \\
        &  0.45--0.50      &  $\rm{1632^{+164}_{-89}   +  1193^{+120}_{-65}  }$    &    0.856        &  0.065        &    $\rm{35.307^{+0.042}_{-0.024}}$ &   $\rm{34.948^{+0.042}_{-0.024}}$ &     $\rm{35.465^{+0.025}_{-0.026}}$   \\
        &  0.50--0.51      &  $\rm{1276^{+1242}_{-358} +   933^{+908}_{-262} }$    &    0.320        &  0.024        &    $\rm{35.094^{+0.295}_{-0.143}}$ &   $\rm{34.735^{+0.295}_{-0.143}}$ &     $\rm{35.252^{+0.169}_{-0.281}}$  \\
\hline
CD Cru  &  0.01--0.09      &  $\rm{1879^{+861}_{-632}  +  1374^{+630}_{-462} }$    &    0.990        &  0.047        &    $\rm{35.798^{+0.164}_{-0.178}}$ &   $\rm{34.875^{+0.164}_{-0.178}}$ &     $\rm{35.847^{+0.133}_{-0.192}}$   \\
        &  0.43--0.51      &  $\rm{1224^{+101}_{-102}  +   895^{+74}_{-75}   }$    &    0.178        &  0.008        &    $\rm{35.425^{+0.034}_{-0.038}}$ &   $\rm{34.502^{+0.034}_{-0.038}}$ &     $\rm{35.474^{+0.031}_{-0.034}}$   \\
        &  0.73--0.83      &  $\rm{1161^{+633}_{-85}   +   849^{+462}_{-62}  }$    &    0.144        &  0.007        &    $\rm{35.380^{+0.189}_{-0.033}}$ &   $\rm{34.457^{+0.189}_{-0.033}}$ &     $\rm{35.428^{+0.107}_{-0.141}}$   \\ 
\hline
\end{tabular}
\end{table*} 